\documentclass[a4paper,11pt]{article}
\pdfoutput=1

\usepackage{jheppub}

\usepackage[utf8]{inputenc}
\usepackage[T1]{fontenc}
\usepackage[english]{babel}
\usepackage{layout}
\usepackage{setspace}
\usepackage{lmodern}
\usepackage{enumitem}
\usepackage{soul}
\usepackage{eurosym}
\usepackage{url}

\usepackage{verbatim}
\usepackage{moreverb}
\usepackage{listings}
\usepackage{fancyhdr}

\usepackage{wrapfig}
\usepackage{graphicx}
\usepackage{pdfpages}

\usepackage{sectsty}

\usepackage{amsthm}
\usepackage{mathrsfs}
\usepackage{cancel}

\usepackage{makeidx}
\usepackage{titlesec}
\usepackage{multirow}
\usepackage{stmaryrd}

\newtheorem{Definition-Proposition}{Definition-Proposition}

\title{Beyond brane-Higgs regularisation: \\ clarifying the method and model}

\author[a]{Andrei Angelescu,}
\author[b]{Ruifeng Leng,}
\author[b]{Gr\'egory Moreau}
\author[b]{and Florian Nortier}

\affiliation[a]{Department of Physics and Astronomy \\ 
University of Nebraska-Lincoln, Lincoln, NE, 68588, USA}
\affiliation[b]{Laboratoire de Physique Th\'eorique \\ 
 CNRS \& Univ. Paris-Sud, Universit\'e Paris-Saclay, 91405 Orsay, France}

\emailAdd{andrei90angelescu@gmail.com}
\emailAdd{ruifeng.leng@u-psud.fr}
\emailAdd{moreau@th.u-psud.fr}
\emailAdd{florian.nortier@th.u-psud.fr}

\abstract{The attractive class of higher-dimensional scenarios, based on a brane-localised Higgs boson coupled to bulk fermions, can address both the puzzle of the structure of the flavour space
and the gauge hierarchy problem. In this framework, a key question arises due to the possibility of fermion wave function discontinuities at the Higgs boundary: how to build rigorously the 
Lagrangian and calculate the fermion mass spectrum as well as the effective four-dimensional (4D) Yukawa couplings? We show that the proper treatment, leading to physically consistent solutions, does not rely on any 
Higgs peak regularisation but requires the presence of certain bilinear brane terms. In particular, no profile jump should appear and the Higgs regularisations turn out to suffer from mathematical discrepancies 
reflected in two non-commutativities of calculation steps debated in the literature. The introduction of bilinear brane terms can alternatively by replaced by vanishing conditions for probability currents at the 
considered flat interval boundaries. Indeed, both contribute to the definition of the field geometrical configuration of the model, even in the free case. The bilinear brane terms could allow to elaborate an ultra-violet  
origin of the chiral nature of the Standard Model and of its chirality distribution among quarks/leptons. The current conditions are implemented through essential boundary conditions to be contrasted with the 
natural boundary conditions derived from the action variation. All these theoretical conclusions are confirmed in particular by the converging exact results of the 4D versus 5D approaches. The analysis is completed by
a description of the appropriate energy cut-off procedure in the present context. The new calculation methods presented, implying the independence of excited fermion masses and 4D Yukawa 
couplings on the `wrong-chirality' Yukawa terms, have impacts on phenomenological results like the relaxing of previously obtained strong bounds on Kaluza-Klein masses induced by 
flavour changing reactions generated via tree-level exchanges of the Higgs field.}

\begin{document} 
\maketitle
\flushbottom

\section{Introduction}

The paradigm of scenarii with extra spatial dimensions (and the composite Higgs models dual via the AdS/CFT correspondance) 
represents an alternative to supersymmetry for addressing the deep gauge {\it hierarchy} problem of the Standard Model (SM). 
In particular, the warped dimension scenarii~\cite{Randall:1999ee} with SM fields in the bulk~\cite{Chang:1999nh}, although relying on a unique 
fundamental energy scale, allow to generate the SM fermion mass {\it hierarchy}~\cite{Gherghetta:2000qt} from a simple geometrical picture of fermion 
profiles (see {\it e.g.} Ref.~\cite{Huber:2001ug,Casagrande:2008hr,Chang:2005ya,Moreau:2005kz,Moreau:2006np,Bouchart:2008vp}).
To realise those two hierarchical features, the Brout-Englert-Higgs scalar field~\cite{Englert:1964et,Higgs:1964ia}, 
providing a mass via the ElectroWeak (EW) symmetry breaking, must be either stuck exactly on the so-called TeV-brane (boundary of the finite extra 
dimension)~\footnote{There exist other phenomenological motivations, like within neutrino mass models, for the Higgs boson to be stuck at the boundary of an 
interval~\cite{Dienes:1998sb,Abada:2006yd,Grossman:1999ra,Huber:2002gp,Moreau:2004qe} or fermions to propagate in the bulk~\cite{Frere:2003hn}.} or located in the 
bulk with a wave function only peaked at the TeV-brane.
In contrast, in the gauge-Higgs unification models, as described for instance in Ref.~\cite{Hall:2001zb}, 
protecting the Higgs mass down to lower energies, the Higgs field propagates all along the extra 
dimension together with matter.

Recently, some attention has been payed on the mathematical context of the interaction between Higgs and fermions both propagating along a 
(warped) extra dimension~\cite{Smolyakov:2011hv}: it was found that to avoid possible pathological behaviours in the fermion sector, constraints on the fermionic field 
Lagrangian must be imposed. Such consistency considerations are interesting from the purely theoretical side and are crucial for the clear understanding of 
higher-dimensional models being now searched and constrained at the Large Hadron Collider (LHC) exploratory phase.

In the present paper, we discuss the rigorous treatment of the other case of a boundary-localised Higgs scalar field, interacting with bulk quark/leptons propagating in a finite interval, 
which presents subtleties that deserve to be looked at more deeply. Such a field configuration occurs in the realistic warped models addressing the 
fermion mass and gauge hierarchy. The case of bulk matter without interactions is also studied.

Let us recall these subtle aspects. First, a question arises about the correct treatment of the specific object that is the Dirac peak entering each 
Lagrangian term which involves the brane-Higgs boson. Secondly, this Dirac peak may induce an 
unusual discontinuity~\footnote{Field jumps may arise in other frameworks~\cite{Bagger:2001qi}.} in the wave function along the 
extra dimension (at the Higgs boundary where further conditions arise from the Lagrangian variations) for some of the bulk fermions: the so-called jump problem~\cite{Csaki:2003sh,Azatov:2009na}. 
These five-dimensional (5D) aspects have motivated the introduction~\cite{Csaki:2003sh,Azatov:2009na} 
of a process of regularisation of the Higgs Dirac peak (smoothing the peak or shifting it from the boundary) in the calculation of Kaluza-Klein (KK) fermion mass 
spectra and effective four-dimensional (4D) Yukawa couplings. Although there is no profound theoretical reason to apply such a regularisation procedure (forcing
interaction-free boundary conditions for fermions), nowadays all the theoretical and phenomenological studies of the warped models with brane-Higgs (see {\it e.g.} 
Ref.~\cite{Casagrande:2008hr,Goertz:2008vr,Casagrande:2010si,Carena:2012fk,Malm:2013jia,Hahn:2013nza}) are relying on this Higgs peak regularisation.

In this paper, we first present the mathematical inconsistencies of this regularisation procedure used in the literature. Then, instead of regularising, 
we develop the rigorous determination of the profiles -- taking into account the mathematical nature of the Dirac peak in the Higgs coupling -- which leads to bulk fermion wave functions without 
discontinuities on the considered extra space. We conclude from this whole approach that neither profile jump nor particular problem arises when a proper mathematical 
framework is used, so that there is in fact no motivation to introduce a brane-Higgs regularisation.

As a consequence, we can now interpret two non-commutativities of calculation steps for Higgs production and decay rates~\cite{Casagrande:2010si,Azatov:2010pf,Carena:2012fk,Malm:2013jia} 
or for fermion masses and 4D Yukawa couplings~\cite{Barcelo:2014kha}, 
previously studied in the literature, to be similar effects and confirmations of the mathematical inconsistencies in the Higgs peak regularisation. 
Besides, the debate in the literature about those two non-commutativities is thus closed by the useless nature of this regularisation.

The correct methods without regularisation, together with their results, are illustrated here in the 
derivation of the KK fermion mass spectrum -- same ideas apply to the calculation of effective 4D Yukawa couplings. This spectrum calculation is done in a simplified 
model with a flat extra dimension, the minimal field content (to write down a Yukawa interaction) and without gauge symmetry. Nevertheless this toy model already possesses all the key ingredients to study 
the delicate brane-Higgs aspects. Hence our conclusions can be directly extended to the realistic warped models with bulk SM matter addressing the fermion 
flavour and gauge hierarchy.

Several new spectrum calculation methods are proposed which further allow confirmations of the analytical results. Those methods go through alternative uses like the 4D or the 5D approach (one extra dimension case), 
and the fermion current determination from the action variations -- we generalise the Noether theorem to include brane-localised terms like the Yukawa couplings -- or by manipulating the equations of motion. 
Besides, the correct derivation of the standard free fermion mass spectrum (in the absence of Yukawa interactions) turns out to be a useful starting guide in particular for the 4D approach or more generically
for a solid comprehension of such higher-dimensional scenarios.

From an historical point of view, the correct method established here arises naturally in the theory of variational calculus as the Lagrangian boundary term 
(brane-Higgs coupling to fermions) is included in a new boundary condition instead of entering the equations of motion~\cite{HilbertBC} (via a regularisation). 
Furthermore the present analysis follows the prescription of considering the Dirac delta to be a distribution. By the way the Dirac peak and distributions were 
formalised and validated mathematically during the 1940's by L.~Schwartz~\cite{Schwartz1,Schwartz2} precisely for the purpose of solving consistently physical problems. Hence today it should 
not be avoided to respect the distribution formalism when facing a physical pro\-blem involving an object like the Dirac delta, as it occurs in the present higher-dimensional context.

The rigorous results obtained for the KK mass spectrum and effective 4D Yukawa couplings are different from the ones derived in general through the Higgs peak regularisation, as it is detailed in the 
present paper. This difference is physical, affecting then phenomenological studies on indirect searches of KK states at high-energy colliders (in particular via the Higgs production and flavour changing neutral currents), and 
analytical (vanishing of the Yukawa coupling with `wrong' fermion chiralities relatively to the SM), which improves the precise theoretical understanding of the higher-dimensional set-up with a brane-localised Higgs field.

Furthermore, the correct mass spectrum obtained here allows to point out the necessity, for bulk fermions (with or without coupling to a brane-localised scalar field), to have certain bilinear brane terms at boundaries 
which are fermion mass terms from the point of view of the spinorial structure but do not introduce new mass parameters~\footnote{The potential 4D effective 
mass involves a dimension-full product of two profile boundary values.}. Indeed, such terms guarantee the existence of physical solutions (with correct profile normalisations, Hermitian conjugate boundary conditions and
satisfying the decoupling limit argument) derived via the least action principle through the variation calculus. Their necessary presence is confirmed by the non-trivial exact matching between the 5D and 4D analytical calculations 
of the mass spectrum.

At a brane without Yukawa coupling, instead of including such a bilinear term, 
we find that one can alternatively impose as an {\it essential boundary condition} (in contrast with {\it natural boundary conditions} coming from the Lagrangian variations) 
the condition of a fermion current along the extra dimension vanishing at this brane -- and exclusively within the 4D approach in case of a brane with localised Yukawa interaction. Indeed, the generic reason for the presence 
of bilinear brane terms is the consistent and complete geometrical definition of models with a finite extra spatial interval in which fermionic matter is stuck. Notice that the choice between the bilinear brane term presence and the 
vanishing fermion current condition relies on the Ultra-Violet (UV) completion of the model. Indeed the vanishing fermion current condition permits alternatively the existence of physical solutions.

Therefore, a first possibility is that the UV completion generates bilinear brane terms for the fermions on both boundaries (those with and without localised Yukawa coupling) of the interval. Then the geometrical interval definition
(interval boundaries and vanishing 5D fermion currents at these boundaries) would be completely contained in the action expression. 
Now in case the UV completion would not induce bilinear brane terms on both boundaries, such essential boundary conditions should be imposed at the brane(s) without bilinear terms
in order to define well the geometrical configuration and to have acceptable physical solutions. We can thus conclude that, whether the geometrical set-up is defined exclusively through the action expression [leading to the natural boundary 
conditions] or (also) via additional essential boundary conditions depends on the origin of the model at high-energies.

In case the UV completion produces bilinear brane terms for the fermions on both boundaries, at low-energies the chiral nature of the SM as well as its field chirality distribution (Left-handed $SU(2)_L$ doublets and Right-handed singlets)
are entirely induced by the signs in front of these bilinear brane terms. This new relation shows how the parti\-cular chiral properties of the SM could be explained by an underlying theory, through the bilinear brane term signs.  
We complete the analysis by a discussion, in this context, on the appropriate treatment of the cut-off in energy due to the framework of higher-dimensional models in a non-renormalisable theory.

The paper is organised as follows. First we describe the minimal model in Section~\ref{A_toy_model_of_flat_extra_dimension}, before presenting the free case and the 4D treatment
of the coupled fermions in Section~\ref{4D perturbative approach}. The 5D approaches are exposed as well, with (Section~\ref{Usual_treatment}) and without 
(Section~\ref{Yukawa_terms_as boundary_conditions}) regularisation. Finally, an overview is provided in Section~\ref{sec:over},
together with a description of the phenomenological impacts. We summarise and conclude in the last section.

\section{The minimal consistent model}
\label{A_toy_model_of_flat_extra_dimension}

\subsection{Space-time structure}

We consider a 5D toy model with a space-time $\mathcal{E}^5 = \mathcal{M}^4 \times \mathcal{C}^1$.
\begin{itemize}
\item $\mathcal{M}^4$ is the usual 4D Minkowski space-time. An event in $\mathcal{M}^4$ is characterised by its 4-vector coordinates $x^\mu$ where 
$\mu =0,1,2,3$ is the Lorentz index. The metric and conventions used are given in Appendix~\ref{notations_and_conventions}. 
\item $\mathcal{C}^1$ is a finite 1D flat compactified extra space. For our purpose we consider the following simple case: 
the interval $\mathcal{C}^1 \equiv [0, L]$, with a length $L \in \mathbb{R}^\star_+$,
parametrised by the continuous extra coordinate $y$ and bounded by two flat 3-branes at $y = 0$ and $y = L$. 
\item A point of the whole 5D space-time $\mathcal{E}^5$ is labeled by its coordinates $z^M$ with an index $M \in \llbracket 0, 4 \rrbracket$. $z^M$ can be split  
into $\left( x^\mu, y \right)$.
\end{itemize}

\subsection{Bulk fermions}

We consider the minimal spin-1/2 fermion field content allowing to write down the 4D effective renormalisable SM Yukawa-like coupling between zero-mode
fermions (of different chiralities) and a scalar field (see Section~\ref{Yukawa}):
a pair of fermions $Q$ and $D$. Both are propagating along the extra dimension, as we have in mind a model extension to a realistic scenario with bulk matter
({\it c.f.} Section~\ref{extension}) where $Q,D$ will be respectively $SU(2)_L$ doublet down-component and singlet quark fields. 
The 5D fields $Q(x^\mu, y)$ and $D(x^\mu, y)$ have thus the following kinetic terms in the covariant 5D action,
\begin{eqnarray}
S_{\Psi} = \int d^4x~dy~ \frac{i}{2} \ (\bar{Q}\Gamma^M \partial_M Q ~-~\partial_M \bar{Q} \Gamma^M Q ~+ \{ Q \leftrightarrow D \} ) \ , \label{eq:actionKin}
\end{eqnarray}
where the last term indicates a field replacement and $\Gamma^M$ denotes the 5D Dirac matrices (see Appendix~\ref{notations_and_conventions}). 
In our notations, the 5D Dirac spinor, being the irreducible representation of the Lorentz group, reads as,
\begin{equation}
Q = \left(\begin{array}{c} Q_L \\ Q_R   \end{array}\right) \;\;\;\; \mbox{and} \;\;\;\;   D = \left(\begin{array}{c} D_L \\ D_R   \end{array}\right)\;,
\label{5DDiracSp}
\end{equation}
in terms of the two two-component Weyl spinors, for the field $Q$ and $D$ respectively. $L/R$ stands for the Left/Right-handed chirality.

Let us rewrite the bulk action of Eq.~\eqref{eq:actionKin} in a convenient form. Using the definition 
$\overleftrightarrow{\partial_M}\hat =\overrightarrow{\partial_M}-\overleftarrow{\partial_M}$ and applying integrations by part along the usual 4-coordinates, it can be recast into $S_{\Psi} = \int d^4x~dy~{\cal L}_{\Psi}$ with
\begin{equation}
{\cal L}_{\Psi} =
\sum_{F = Q, D} \left\{ i F^\dagger_R \sigma^\mu \partial_\mu F_R + i F^\dagger_L \bar{\sigma}^\mu \partial_\mu F_L + \dfrac{1}{2} \left( F^\dagger_R \overleftrightarrow{\partial_4} F_L 
- F^\dagger_L \overleftrightarrow{\partial_4} F_R \right) \right\}  \ ,
\label{L_3}
\end{equation}
omitting the global 4-divergence which must vanish in the action integration due to vanishing fields at (infinite) boundaries. 
Indeed, when minimising the action, we see that the varied terms must vanish separately at (infinite) boundaries, since the generic non-vanishing field variations at boundaries are independent from each other and from bulk ones. 
This is realised by the standard configuration of vanishing fields themselves at boundaries which is compatible with the wave function normalisation conditions.

\subsection{Bilinear brane terms}
\label{brTermSec}

Interestingly, in the absence of vanishing fermion current condition at a boundary of the considered interval $[0, L]$, 
the presence at this 3-brane of some bilinear terms, for bulk fermions being either free or coupled to a scalar field on this brane, turns out to be necessary.  
Indeed, these bilinear terms insure the existence of physical solutions [see Section~\ref{4D perturbative approach} for the 4D approach and Section~\ref{Yukawa_terms_as boundary_conditions} for the 5D one] deduced from 
the least action principle. The theoretical reason for the presence of the Bilinear Brane Terms (BBT) at the boundaries of the interval is the correct geometrical configuration definition for models where fermions cannot propagate 
beyond the two boundaries, as will also be described in Sections~\ref{4D perturbative approach} and \ref{Yukawa_terms_as boundary_conditions}. These sections will also point out the 4D/5D matching of the mass spectrum
exact result which constitutes in particular a confirmation for the necessary presence and exact form (including coefficients) of the BBT.

Necessary BBT read as~\footnote{Similar terms, 
leading in particular to $\mathcal{L}_{B}=\dfrac{1}{2}(\bar{Q}^D Q^D-\bar{D}D)$, would hold in a model version extended to the EW symmetry of the SM, with the $Q$ field promoted to an $SU(2)_L$ doublet. 
In contrast, terms of the kind $\bar{Q}^U D$ (or $\bar{Q}D$), $\bar{Q}^U Q^D$ or $\bar U D$ would obviously not belong to a gauge invariant form.}, 
\begin{eqnarray}
S_{\rm{B}} = \int_0^L d^4x~dy~ \left [ \ \left \{ \delta(y-L) - \delta(y) \right \} \ \mathcal{L}_{B} \ \right ]  \ , \ \mbox{with} \ \ \mathcal{L}_{B}=  \dfrac{1}{2} \left(  \bar{Q}Q - \bar{D}D \right) \ , \label{eq:actionBound}
\end{eqnarray}
where $\delta(y-L)$ denotes the Dirac peak at $y=L$. Indeed, those BBT will lead to the set of boundary conditions in Eq.~\eqref{completeEBCsm0} for the wave functions $q^n(y),d^n(y)$ of the 5D fields $Q,D$ which then 
possess a non-vanishing normalisable zero-mode ($m_{[n=0]}=0$) for only one chirality [$L$ or $R$ as $\sin(m_{[n=0]}\, y)=0$]; hence at low-energies (below the first KK mass eigenvalue $m_1$), only one chirality of a given 4D field arises 
in the KK decomposition~\eqref{KK_1} so that one recovers the chiral nature of the SM. 
Furthermore, within an extended realistic model (as described in Section~\ref{extension}) where the $Q^{(D)}$ field would be the down-component of an SM $SU(2)_L$ gauge doublet, the unique chiralities of the zero-mode
4D fields $Q_L^{(D)\, 0}(x_\mu)$ and $D_R^0(x_\mu)$ predicted by Eq.~\eqref{completeEBCsm0} via Eq.~\eqref{KK_1} would well correspond to the SM chirality configuration~\footnote{Taking 
the opposite sign for each of the four terms in Eq.~\eqref{eq:actionBound} would lead to exchanged boundary conditions between $q^n(y)$ and $d^n(y)$ relatively to Eq.~\eqref{completeEBCsm0} and in turn to another 
chirality configuration.}. Notice that Eq.~\eqref{KK_1} [involving KK modes rather than mass eigenstates] and Eq.~\eqref{completeEBCsm0} are valid within the relevant 4D treatment of the localised 
Yukawa interaction where it is explicit that the SM particles (whose mass mainly originates from the EW symmetry breaking) are indeed mainly composed by the zero-modes (small mixings with the massive KK states), 
as imposed by the small experimental deviations generally observed with respect to the theoretical SM predictions.

Therefore, it is remarkable that
the BBT allow to make a step towards the UV explanation of the well-known SM chiral properties (chiral nature and chirality configuration) by directly linking these chiral aspects to explicit signs in front of Lagrangian terms (BBT signs), 
as described right above. Then the last step would be to build a UV completion of the model to generate these BBT signs.
In other words, the entire control of the chiral structure by the BBT signs is a new feature that shows how an underlying theory could produce the SM chiral structure.

For completeness, we mention that the two other BBT sign configurations, 
\begin{eqnarray}
S'_{\rm{B}} = \int_0^L d^4x~dy~ \left [ \ \left \{  \delta(y-L) + \delta(y)  \right \} \ \mathcal{L}'_{B} \ \right ]  \ , \ \mbox{with} \ \ \mathcal{L}'_{B}= \sigma \,  \dfrac{1}{2} \ \bar{F}F  \label{eq:actionBoundCUSTO}
\end{eqnarray}
and $\sigma = \pm 1$, for 5D fields of the form~\eqref{KK_1} lead to the two sets~\eqref{completeEBCcusto0} of boundary conditions
and in turn to a vector-like field content, as for the so-called custodian fermions in custodially protected warped models~\cite{Agashe:2003zs}.
Indeed, Eq.~\eqref{completeEBCcusto0} leads to the absence of zero-modes ($m_{[n=0]}\neq 0$) and hence any KK state has both Left and Right chiralities.
Notice here that Eq.~\eqref{KK_1} and Eq.~\eqref{completeEBCcusto0} are valid for the free case. 
Once again, the control of the vectorial structure by the BBT signs is a novel characteristic that shows how a UV completion could produce a vector-like field content.

What is the direct effect of the BBT~\eqref{eq:actionBound} on the final fermion mass eigenvalues?
In the 4D approach and in the case without Yukawa interaction [see Section~\ref{4D perturbative approach}], these BBT have no effect on the 4D fermion mass matrix [Eq.~\eqref{M}]: 
after injecting the profile solutions, those BBT vanish due to the induced boundary conditions of Eq.~\eqref{completeEBCsm0}
which impose that one of the two wave functions ($L$ or $R$)~\footnote{For instance, $\bar D D = D^\dagger_L D_R + D^\dagger_R D_L$.} entering the BBT 5D fields [{\it c.f.} Eq.~\eqref{KK_1}] is  
equal to zero, at $y=0$ [$\sin(m_{n}\, 0)=0$] and $y=L$ [$\sin(m_{n}\, L)=0$], systematically for each one of the two Lagrangian BBT~\eqref{eq:actionBound}. In contrast,
in the 5D approach, the BBT~\eqref{eq:actionBound} play a numerical and direct r\^ole in the fermion mass spectrum, through the boundary conditions coming from the action variations 
[see Section~\ref{Yukawa_terms_as boundary_conditions}].

Formerly, this kind of bilinear fermion brane terms~\eqref{eq:actionBound}-\eqref{eq:actionBoundCUSTO} was first introduced by hand to derive the more specific AdS/CFT correspondance 
in the calculation of correlation functions for spinors~\cite{Henningson:1998cd,Mueck:1998iz} -- the exact AdS/CFT duality being possibly realised in the UV completion of warped models 
(from which the present simplified scenario is inspired). Then within this AdS/CFT paradigm, similar boundary terms have been added at the UV-brane only ($y=0$) 
to guarantee the minimisation of the action in the holographic version of the warped model with bulk fermions~\cite{Contino:2004vy}. The least action principle was also invoked 
in Ref.~\cite{Henneaux:1998ch} to justify such bilinear fermion brane terms in the AdS/CFT context and through the path integral formalism. Equivalently, still in the AdS/CFT framework, 
these terms have been motivated in the Lagrangian density from an action form involving explicitly the Hamiltonian (to obtain a consistent Hamiltonian formulation 
when perfor\-ming the Legendre transformation)~\cite{Arutyunov:1998ve}. 
Other boundary-localised terms were also introduced in a field theory defined on a manifold with boundaries within the context of gravity: the Gibbons-Hawking boundary 
terms~\cite{Gibbons:1976ue, Chamblin:1999ya, Lalak:2001fd, Carena:2005gq}. Those terms are needed to cancel the variation of the Ricci tensor at the boundaries of the manifold.

The finite geometry set-up is defined via either the BBT inclusion or the vanishing fermion current condition, depending on the considered UV completion of the model. From the point of view of the effective field theory, 
it means that it can happen that the underlying theory does not forbid (through a short-distance mechanism or a residual symmetry) any possible non-renormalisable Lorentz-invariant operator involving the 
5D fields $Q,D$ (including covariant derivatives) up to dimension 5 -- this dimension choice being motivated in Section~\ref{Yukawa} -- in the low-energy effective model described in 
this Section~\ref{A_toy_model_of_flat_extra_dimension}. 
Then the present fermionic operators would be those included in the considered Actions~\eqref{L_3} (dimension 5 operators) and \eqref{eq:actionBound} (dimension 4 operators): the BBT part.

Notice that bulk mass terms, usually modifying the bulk fermion profiles, bring useless complications so we will not 
consider them in our present calculations, as the paper conclusions on fermion couplings to a brane-field can be easily extended~\cite{WIP}.

\subsection{Brane-localised scalar field}

The subtle aspects arise when the fermions couple to a single 
4D scalar field, $H$, confined on a boundary taken here to be at $y=L$ (as inspired from warped scenarii addressing the gauge hierarchy
problem). The action of this scalar field has the generic form
\begin{eqnarray}
S_H  = \int d^4x~dy~ \delta(y-L) \ {\cal L}_{\rm{H}} \ , \ \mbox{with} \ \ {\cal L}_{\rm{H}} =   (\partial_\mu H)^\star (\partial^\mu H) - V(H)     
\label{eq:actionH}
\end{eqnarray}
where the potential $V$ possesses a minimum which generates a non-vanishing 
Vacuum Expectation Value (VEV) for the field developed as $$H=\frac{v+h(x^\mu)}{\sqrt{2}}$$ in analogy with the SM Higgs boson.

Note that one could think of replacing (up to a constant) the $\delta(y-L)$ peak in Eq.~\eqref{eq:actionH} by an Heaviside step~\footnote{$\Theta(r)=0$ for $r>0$ and $\Theta(r)\hat =1$ for $r=0$.} function $\Theta(L-y)$ 
that could play a similar r\^ole of localising the scalar field Lagrangian at the boundary $y=L$. Nevertheless, the integration in Eq.~\eqref{eq:actionH} over 
the interval $[0, L]$ would then be strictly equivalent to the integration over $[0, L[$   
%GMF
and in turn equal to zero given the vanishing Heaviside function value there. Such a situation would in fact correspond to the total 
absence of the $H$ scalar field which conflicts with the considered field content hypothesis.

\subsection{Yukawa-like interactions}
\label{Yukawa}

We focus on the following basic interaction in order to study the subtleties induced by the brane-scalar field coupling to bulk fermions,
\begin{eqnarray}
S_{\rm{Y}} = - \int d^4x~dy~ \delta(y-L) \ {\cal L}_{\rm{Y}} \ , \ \mbox{with} \ \ {\cal L}_{\rm{Y}} 
= Y_5\ {Q}^\dagger_LHD_R ~+~ Y^\prime_5\ {Q}^\dagger_RHD_L ~+~  {\rm H.c.} \ \  \label{eq:actionYuk}
\end{eqnarray}
Considering operators, involving $H$, $Q$ and $D$, up to dimension 5 allows to include this Yukawa coupling of interest~\footnote{Notice that for instance a dimension-6
operator of type $\frac{1}{M^2}\delta(y-L){Q}^\dagger_{L/R}H^2D_{R/L}$, $M$ being a mass scale, would be treated in a similar way as the couplings in Eq.~\eqref{eq:actionYuk} 
(and can contribute to the Yukawa couplings~\eqref{eq:actionYuk} through the scalar field VEV).}.
The coupling constants $Y_5$ and $Y^\prime_5$ of Yukawa type, entering these two distinct terms, are independent [{\it i.e.} parameters with possibly different values]  
as a well-defined 4D chirality holds for the fermion fields on the 3-brane strictly at $y=L$ (see for instance Ref.~\cite{Azatov:2009na}).

In order to avoid the introduction of a new energy scale, one could define the 5D Yukawa coupling constants by giving their explicit dependence in $L$: $Y_5=Y_4L$ and $Y^\prime_5=Y_4'L$, 
where $Y_4,Y_4'$ are dimensionless coupling constants of ${\cal O}(1)$. Then $Y_4$ can be identified with the SM Yukawa coupling constant,
as shown when applying the decoupling limit (infinitely heavy KK masses and any new physics energy scale)~\footnote{Note that in the decoupling limit where in particular $L \to 0$, generally $Y_5 \to 0$ due to the 
dimension of the 5D Yukawa coupling constants.}.

From now on, we restrict our considerations to the VEV of $H$ as the aim is to calculate the KK fermion mass spectrum which
is unaffected by the interactions of the $h(x^\mu)$ fluctuation field with fermions. Hence, we concentrate on the following action issued from Eq.~\eqref{eq:actionYuk},
\begin{equation}
S_X = - \int d^4x~dy~ \delta(y-L) \ {\cal L}_{\rm{X}} \ , \ \mbox{with} \ \ {\cal L}_{\rm{X}} 
= X\ {Q}^\dagger_L D_R ~+~ X^\prime\ {Q}^\dagger_RD_L ~+~  {\rm H.c.} \  \
\label{L_5}
\end{equation}
with the compact notations $X = \dfrac{v Y_5}{\sqrt{2}}$ and $X' = \dfrac{v Y_5'}{\sqrt{2}}$.

\subsection{Model extension}
\label{extension}

The toy model considered is thus characterised by the Lagrangian
\begin{eqnarray}
S_{\rm{5D}} = S_{\Psi} + S_B + S_H  + S_X \ . \label{eq:actionTot}
\end{eqnarray}
Nevertheless, the conclusions of the present paper can be directly generalised to realistic warped models with bulk SM matter 
solving the fermion mass and gauge hierarchies. Indeed, working with a warp extra dimension instead of a flat one would not affect the conceptual subtleties about 
coupling bulk fermions to a brane-scalar field~\cite{WIP}. The boundaries at $y=0$ and $y=L$ could then become respectively the Planck and TeV-branes. 
Similarly, the scalar potential, $V(H)$, can be extended to any potential [like the SM Higgs potential breaking the EW symmetry] as long as it still generates a 
VEV for the scalar field as here. In this context the $H$ singlet can be promoted to the Higgs 
doublet under the SM $SU(2)_L$ gauge group, simply by inserting doublets 
in the kinetic term of Eq.~\eqref{eq:actionH}. The whole structure of the coupling of Eq.~\eqref{L_5} between bulk fermions and the localised VEV would as well
remain identical in case of fermions promoted to SM $SU(2)_L$ doublets: after group contraction of the doublet $(Q^U,Q^D)^t$ with down/up-quark 
singlets $D,U$, one would obtain two replica of the structure~\eqref{L_5} with the forms $Q_C^{D\dagger} D_{C'}$ and $Q_C^{U\dagger} U_{C'}$ where $C^{(\prime)}\equiv L,R$ denotes 
the chirality. Hence the procedure described in this paper should just be applied to both terms 
separately~\footnote{The fermion actions in Eq.~\eqref{L_3} and \eqref{eq:actionBound} would be trivially generalised as well to a scenario with a gauge symmetry.}. 
A same comment holds for the SM colour triplet contraction 
and the field content extension to the three flavours of quarks and leptons of the SM. Notice that the flavour mixings would be combined with the mixings among
fermion modes of the KK towers, without any impact on the present considerations about brane-localised couplings.

\section{4D Perturbative approach}
\label{4D perturbative approach}

\subsection{5D aspects for (formally) free fermions}
\label{free_bulk_fermions}

In this part, we calculate the fermionic mass spectrum in the basic case where $Y_5=Y_5'=0$ in Eq.~\eqref{eq:actionYuk} (studied in various references~\cite{Cheng:2010pt, Csaki:2003sh, Contino:2004vy, Csaki:2005vy, Gherghetta:2006ha, Grojean:2007zz, Dobrescu:2008zz, Gherghetta:2010cj, Ponton:2012bi, Barcelo:2014kha, Raychaudhuri:2016}), pointing out the correct treatment. Let us also remark that in this case there is no 
4D/5D matching condition to look at (pure 5D calculation of the masses). The main interest of this section is to develop the rigorous procedure for applying the boundary conditions. 

\subsubsection{Absence of Yukawa couplings}
\label{no_Yukawa}

\paragraph{\textit{Natural boundary conditions}} $\phantom{0}$ \vspace{0.3cm} \\
In order to extract from the relevant Lagrangian~\eqref{L_3} the Equations Of Motion (EOM) and the Boundary Conditions (BC) for the bulk fermions, we apply the 
least action principle -- or Hamilton's variational principle -- for each of them ($F=Q,D$~\footnote{EOM and BC for the fields and their conjugates are trivially related through Hermitian conjugation.}).
Assuming, at a first level, the boundary fields $F(x^\mu, y=\{0,L\})$ $\hat = \left . F \right |_{0, L}$ to be initially unknown (unfixed), they 
should be deduced from the action minimisation with respect to them, considering thus non-vanishing generic~\footnote{A field 
%GMF
variation reads as $\delta F(z^M)=\epsilon\,  \eta (z^M)$ with a generic function $\eta (z^M)$ and an infinitesimal parameter $\epsilon \to 0$.}
variations $\left . \delta F \right |_{0, L}  \neq 0$~\footnote{Then in the final step, once for instance the field $\left . F \right |_{L}$ is found and fixed by the solution (not initially fixed as an hypothesis in this considered case),
its resulting determined 
form does not imply $\left . \delta F \right |_{L}  = 0$ which would be incompatible with the starting non-vanishing field variation: there are sometimes confusions in articles about these chronological aspects of the variational calculus.}. 
In other words, $\left . F \right |_{0, L}$ should be then obtained from the so-called
Natural Boundary Conditions (NBC). The stationary action condition can be split, without loss of generality (the functional variations are generic so that $\delta{\bar{Q}}$ and $\delta{\bar{D}}$ 
are independent), into action variations with respect 
to each field $\bar Q$ and $\bar D$: 
\begin{equation}
0 = \delta_{\bar{F}} S_{\Psi} = \displaystyle{ \int d^4x \; \int_0^{L} dy \;  \left[ \delta \bar{F} \; i \Gamma^M \partial_M F + \dfrac{1}{2} \ \partial_4 \left( \delta \bar{F} \; \gamma^5 F \right) \right] }
\label{HVP_1}
\end{equation}
as written after we have expressed Eq.~\eqref{L_3} in terms of the four-component spinors, based on the Dirac matrices of Appendix~\ref{notations_and_conventions}, and integrated by part
its last two terms.
Note that $\delta (\partial_M \bar{F}^\alpha) \partial \mathcal{L}_{\Psi}/\partial ( \partial_M \bar{F}^\alpha ) =0$ [$\alpha$ being the implicitly summed 
spinor index] was used to obtain Eq.~\eqref{HVP_1}. 
Besides, the variations $\delta \bar F |_{0, L} \neq 0$ appear in the second term of this equation (pure boundary terms after integration over $y$ of the global 
$\partial_4$ derivative) generated by the action minimisation with respect to $\bar F$ at the boundaries.

Following the theory of variational calculus~\cite{Giaquinta,Burgess:2002ea}, 
the distinct terms of Eq.~\eqref{HVP_1} vanish separately (respectively the volume and the surface terms)
to insure that $\delta_{\bar{F}} S_{\Psi} = 0$ still for generic and in turn independent field variations: 
\begin{equation}
\delta_{\bar{F}} S_{\Psi} = 0 \ \Rightarrow \ 
\left\{
\begin{array}{c c  l}
 0 &=& \displaystyle{ \delta \bar{F} \; i \Gamma^M \partial_M F } \ , \ \ \forall \, x^\mu, \ \forall \, y \in [0,L]
\\ \vspace{-0.3cm} \\
0 &=& \displaystyle{ \delta \bar{F} \; \gamma^5 F |_{0} =  \delta \bar{F} \; \gamma^5 F |_{L} }  \ , \ \ \forall \, x^\mu
\end{array}
\right.
\label{HVP_2}
\end{equation}
where the first (second) line constitutes the bulk EOM (NBC)~\footnote{We find the Hermitian conjugate EOM and NBC by integrating by part 
the bulk piece of the relation $\delta_{F} S_{\Psi} = 0$ (non-vanishing boundary terms from integration over the extra dimension then contribute) in order to get rid of the field factors $\partial_M \delta F$.}. 
Notice that the NBC origi\-nate from the last term of Eq.~\eqref{HVP_1} obtained after an integration by part of the initial action.  
Using once more the fact that for the searched bulk fields, the $\delta \bar{F}^\alpha (z^M)$ for any $\delta \bar{F}^\alpha$
$[\alpha=0,1,...,4]$ are independent from each other and non-vanishing, it is useful for the following to recast Eq.~\eqref{HVP_2} into these two-component spinor relations
(still using Appendix~\ref{notations_and_conventions}),  
\begin{equation}
 i \Gamma^M \partial_M F  = 0 \  \Leftrightarrow \ 
\left [ 
\begin{array}{r c l}  
i \sigma^\mu \partial_\mu F_R   &=& - \partial_4 F_L  
\\ \vspace{-0.2cm} \\
i \bar{\sigma}^\mu \partial_\mu F_L  &=& \partial_4 F_R  
\end{array}
\right.
\label{ELE_1}
\end{equation}
and,
\begin{equation}
\gamma^5 F |_{0} = \gamma^5 F |_{L} = 0 \  \Leftrightarrow \ F_L \vert_{0} = F_R \vert_{0} = F_L \vert_{L} = F_R \vert_{L} = 0 \ .
\label{BC_1}
\end{equation}
Let us now deduce, from those equations involving the 5D fields, the relations on their profiles along the extra dimension.

\paragraph{\textit{The naive approach}} $\phantom{0}$ \vspace{0.3cm} \\
To develop a 4D effective picture, let us replace the 5D fields, in the relations obtained just above, by their standard solution in the form of a KK decomposition, 
\begin{equation}
F_{L/R} \left( x_\mu, y \right) = \dfrac{1}{\sqrt{L}} \displaystyle{ \sum_{n=0}^{+\infty} f^n_{L/R}(y) \; F^n_{L/R} \left( x_\mu \right)} \ ,
\label{KK_1}
\end{equation}
where $f^n_{L/R}=q_{L/R}^n$ or $d_{L/R}^n$ are the dimensionless wave functions along the extra dimension associated respectively to the 
4D fields $F^n_{L/R}=Q_{L/R}^n$ or $D_{L/R}^n$ of the KK excitations. The integer $n$ is defined throughout the whole paper as being the level index of the heavy fermion mode [here the KK state~\footnote{Not yet 
the mass eigenstate in case of Yukawa interactions.}] 
tower and is chosen to be positive; the meaningful feature about the general KK decomposition~\eqref{KK_1}  is rather the infinite summation (possibly also from $-\infty$ to $+\infty$) dictated by 
field expressions as Fourier series on a finite interval.

Inserting Eq.~\eqref{KK_1} into the 5D field relations~\eqref{ELE_1} and using the following two-component Weyl equations for the 4D fermions 
(issued from the four-component Dirac equation)~\footnote{Within the natural unit system.}, 
\begin{equation}
\left\{
\begin{array}{r c l}
i \bar{\sigma}^\mu \partial_\mu F_L^n \left( x_\mu \right) &=& m_n \; F_R^n \left( x_\mu \right)
\\ \vspace{-0.2cm} \\
i \sigma^\mu \partial_\mu F_R^n \left( x_\mu \right) &=& m_n \; F_L^n \left( x_\mu \right)
\end{array}
\right. \ ,
\label{Dirac_1}
\end{equation}
where $m_n$ are the KK masses~\footnote{Also mass eigenvalues in absence of Yukawa interactions.} for the fermions~\footnote{The same masses $m_n$ 
enter the Weyl equations for the $Q_{L/R}^n (x_\mu)$ and $D_{L/R}^n (x_\mu)$ fields which are des\-cribed by separate and identical terms in the considered Lagrangian.}, one can directly extract 
these differential free equations for the profiles:
\begin{equation}
\forall n \geq 0 \, , \ 
\left\{
\begin{array}{r c l}
\partial_4 q_R^n(y) &=& m_n \; q_L^n(y)
\\ \vspace{-0.2cm} \\
\partial_4 q_L^n(y) &=& - m_n \; q_R^n(y)
\\ \vspace{-0.2cm} \\
\partial_4 d_R^n(y) &=& m_n \; d_L^n(y)
\\ \vspace{-0.2cm} \\
\partial_4 d_L^n(y) &=& - m_n \; d_R^n(y)
\end{array}
\right. \ .
\label{ELE_2}
\end{equation}
These four equations have been obtained by writing the equality, per KK level, between each term of the KK state sums on the left-hand side and right-hand side of Eq.~\eqref{ELE_1} 
(and by simplifying thanks to identical 4D fields on each side), instead of considering compensations between different terms in Eq.~\eqref{ELE_1} which would mean having physical 
4D fields $F_{L/R}^n \left( x_\mu \right)$ [solutions of Eq.~\eqref{Dirac_1}] expressed as linear combinations of other mass eigenstates $F_{L/R}^{n'} \left( x_\mu \right)$: 
such re-expressions would induce, in the Lagrangian, KK mass mixing terms for the mass eigenstates which is not consistent.

Deriving and combining the first order differential equations~\eqref{ELE_2}, one can decouple them into the second order differential equations
\begin{equation}
\forall n \geq 0 \, , \  \partial_4^2 f_{L/R}^n(y) = - m_n^2 \; f_{L/R}^n(y)
\label{ELE_3}
\end{equation}
being the equations of independent harmonic oscillators, whose solutions possess the general form 
\begin{equation}
\forall n \geq 0\, , \ f_{L/R}^n(y) = A^n_{L/R} \; \cos(m_n \; y) + B^n_{L/R} \; \sin(m_n \; y) 
\label{profiles_0}
\end{equation}
where $A^n_{L/R}$, $B^n_{L/R}$ are constant parameters.

Now inserting Eq.~\eqref{KK_1} into the 5D field conditions of Eq.~\eqref{BC_1}, we obtain the following vanishing conditions for any profile
\begin{equation}
\forall n \geq 0~\footnote{Throughout this paper, the notation ``$\forall n  \geq 0$'' applies on the natural integer $n$ defined in Eq.~\eqref{KK_1}: $n=0,1,2,3,\dots$}
\, , \ f_{L/R}^n(0) =  f_{L/R}^n(L) = 0 
\label{profiles_vanish}
\end{equation}
avoiding inconsistent relations among mass eigenstate 4D fields, as explained below Eq.~\eqref{ELE_2}.

Those conditions, combined with Eq.~\eqref{ELE_2} and Eq.~\eqref{profiles_0}, give rise to the vanishing profiles $f_{L/R}^n(y)=0$ 
($\forall n \geq 0$)~\footnote{$f_L^n(0)=0$, $\forall n$, and Eq.~\eqref{profiles_0} 
lead to $A^n_{L}=0$ so that $f_{L}^n(y) = B^n_{L} \, \sin(m_n \, y)$. Then Eq.~\eqref{ELE_2} induces $f_{R}^n(y) = A^n_{R} \, \cos(m_n \, y)$ and in turn $f_{R}^n(y) = 0$
from $f_R^n(0)=0$. Hence Eq.~\eqref{ELE_2} imposes $B^n_{L}=0=f_{L}^n(y)$.} 
which are not compatible with the two ortho-normalisation conditions for $f^n_{L}(y)$ and $f^n_{R}(y)$,
\begin{equation}
\dfrac{1}{L} \int_0^{L} dy \; f^{n\star}_{L/R}(y) \; f^m_{L/R}(y) = \delta_{nm} \ , \forall n  \geq 0 \ , \forall m \geq 0 \ ,
\label{normalization_1}
\end{equation}
coming out from the imposition of canonical and diagonal normalised kinetic terms for the 4D fields after inserting the KK decomposition~\eqref{KK_1} into the 5D field
kinetic terms~\eqref{eq:actionKin}.

The theoretical inconsistency obtained here for the considered free model reveals a problem in the treatment of a simple boundary without localised couplings to bulk matter
(which is the case of both boundaries here). The correct treatments, based on either fermion current conditions at the boundaries
or boundary-localised terms (the BBT), are exposed respectively in the two following subsections.

\subsubsection{Introducing the fermion current condition [EBC]}
\label{sec:freeEBC}

\paragraph{\textit{The current from action variations}} $\phantom{0}$ \vspace{0.3cm} \\
In fact, the free version [$Y_5=Y_5'=0$] of the model defined in Section~\ref{A_toy_model_of_flat_extra_dimension} (and finite extra dimension scenarii in general) 
does impose conditions on the bulk fermions at the extra dimension boundaries, which were not included in the above naive analysis. 
These conditions contribute to define the geometrical field configuration of the considered model. They will constitute the so-called Essential 
Boundary Conditions (EBC), as imposed by the model definition, which are complementary to the NBC already defined in Eq.~\eqref{HVP_2}. Indeed the NBC come from an integration by part of the initial action 
with respect to the fifth dimension over the interval $[0,L]$ and thus take into account the space-time structure itself.

Regarding the geometrical field configuration within the present free model, 
each fermion field is defined only along the interval $[0,L]$. This model building hypothesis, that fermions neither propagate towards nor come from the
outside of a finite range, translates into the condition of vanishing probability current at both boundaries for each independent fermion species separately (without possible compensations).

Formally speaking, after having varied the Lagrangian~\eqref{eq:actionKin} [see Eq.~\eqref{HVP_1}] and in turn derived the bulk EOM~\eqref{ELE_1} [from the first relation of Eq.~\eqref{HVP_2}]
as well as the NBC [second relation of Eq.~\eqref{HVP_2}], the application of the Noether theorem demonstrated in Appendix~\ref{Noether_THM} (by using the EOM)~\footnote{Valid trivially in the absence of BBT as well.}
gives rise to the two probability currents~\eqref{currentGEN} defined independently for the two bulk fermions~\footnote{See
Ref.~\cite{Cheng:2010pt} for scalar field currents.} represented by the 5D fields $F=Q,D$:
\begin{equation}
j^M_Q = - \alpha \bar{Q} \; \Gamma^M Q \ , \  j^M_D = - \alpha' \bar{D} \; \Gamma^M D \phantom{0} \, ,
\label{courant_1}
\end{equation}
associated to the two global $U(1)_F$ symmetries of the free Action~\eqref{eq:actionKin} corresponding respectively to the distinct transformations,
\begin{equation}
Q \mapsto \text{e}^{i \alpha} Q \ , \ D \mapsto \text{e}^{i \alpha'} D  \ .
\label{symetrie_1} 
\end{equation}
$\alpha, \alpha'$ ($\in \mathbb{R}$) are continuous parameters entering for instance the infinitesimal field varia\-tions~\footnote{We 
use different notations for the infinitesimal field variations under specific transformations, $\underline{\delta} F$, and generic field variations in the variation calculus, $\delta F$.}:
\begin{equation}
\underline{\delta} Q = i \alpha Q \, , \  \underline{\delta} \bar Q = - i \alpha \bar Q \  .
\label{delta-bar-free} 
\end{equation}

Now the 4 conditions of vanishing probability currents are thus,
\begin{equation}
\left. j_F^4 \right |_{0, L}= - \alpha^{(\prime)} \left. \bar{F} \; \Gamma^4 F \right |_{0, L} = i \alpha^{(\prime)} \left. \left( F^\dagger_R F_L - F^\dagger_L F_R \right) \right|_{0, L} = 0 \, , \ \forall x^\mu \ ,
\label{courant_2}
\end{equation}
where we have used Eq.~\eqref{5DDiracSp}. For a non-trivial transformation~\eqref{symetrie_1} [$\alpha^{(\prime)}\neq 0$],
the condition~\eqref{courant_2} on the 4D fields $F^n_{L/R} \left( x_\mu \right)$ [{\it c.f.} Eq.~\eqref{KK_1}] entirely fixed by Eq.~\eqref{Dirac_1} leads to an equation
on the 4D space-time coordinates and momenta whereas the fields of the considered model 
must be defined in the whole 4D space-time and for any 4D momentum. The most general way out is to make of Eq.~\eqref{courant_2} a trivial equality by having
\begin{equation}
\left. F_L \right |_{0}=0 \  \text{or} \ \left. F_R \right |_{0}=0 , \phantom{0} \text{and,} \phantom{0}   \left. F_L \right |_{L}=0 \  \text{or} \ \left. F_R \right |_{L}=0
\label{courant_3}
\end{equation}
corresponding to a vanishing coefficient in each term of the condition~\eqref{courant_2}:
\begin{equation}
f^n_L(0)=0 \  \text{or} \ f^n_R(0)=0 , \phantom{0} \text{and,} \phantom{0}  f^n_L(L)=0 \  \text{or} \ f^n_R(L)=0 , \forall n \geq 0,
\label{courant_4}
\end{equation}
which avoids inconsistent relations among mass eigenstate 4D fields, as discussed below Eq.~\eqref{ELE_2}.

These necessary conditions~\eqref{courant_3} of vanishing fields at boun\-daries are the EBC and correspond to some fields initially fixed at boun\-daries. 
Having such known fields at boun\-daries imposes~\cite{Giaquinta} to have vanishing functional variations, 
\begin{equation}
\left . \delta F_L \right |_{0}  = 0 \  \text{or} \ \left . \delta F_R \right |_{0}  = 0 , \phantom{0} \text{and,} \phantom{0}  \left . \delta F_L \right |_{L}  = 0 \  \text{or} \ \left . \delta F_R \right |_{L}  = 0 \ .
\label{delta-BC}
\end{equation}
There is an overall consistency since no action minimisation with respect to a field 
$\left . F_{L/R} \right |_{0, L}$ (relying on $\left . \delta F_{L/R} \right |_{0, L}\neq 0$) is needed for such a known fermion field at a boundary. 
In contrast with the first treatment [above] where the boundary fields $\left . F \right |_{0, L}$ were assumed to be initially unknown  
and then found out (NBC) through the least action principle.

The new EBC~\eqref{courant_3} must be combined with the obtained NBC [second relation of Eq.~\eqref{HVP_2}] which read as,
\begin{equation}
\left. \left( \delta F_L^\dagger F_R \right) \right|_{0} - \left. \left( \delta F_R^\dagger F_L \right) \right|_{0} = \left. \left( \delta F_L^\dagger F_R \right) \right|_{L} - \left. \left( \delta F_R^\dagger F_L \right) \right|_{L} = 0 \ .
\label{complete_boundary_term}
\end{equation}
In fact each of the four sets of EBC in Eq.~\eqref{courant_3}-\eqref{delta-BC} imply the NBC~\eqref{complete_boundary_term} so that it is sufficient to consider these EBC.

In other words, when deriving the NBC, before knowing the EBC (as described above), one would sum generically in the action variations~\eqref{HVP_1} the terms with all non-vanishing field variations at boundaries but, 
once the EBC~\eqref{courant_3} are determined and selected (fixing some fields at boundaries accordingly to Eq.~\eqref{delta-BC}), keep explicitly only the known non-vanishing variations ({\it i.e.} omit to vary the action with respect to 
known fields), do not consider null terms and in turn similarly in the NBC (so that some terms of Eq.~\eqref{complete_boundary_term} are omitted). Then the resulting NBC and EBC can be combined.

Now the general solutions~\eqref{profiles_0} of the decoupled equations derived from the EOM~\eqref{ELE_1}, once re-injected into the initial equations~\eqref{ELE_2} on the profiles, become 
\begin{equation}
f_{L}^n(y) = B^n_{R} \; \cos(m_n \; y) - B^n_{L} \; \sin(m_n \; y), \ \ 
f_{R}^n(y) = B^n_{L} \; \cos(m_n \; y) + B^n_{R} \; \sin(m_n \; y).
\label{profiles_rein}
\end{equation}
These solutions are taken continuous at the boundaries in order to possibly have well-defined derivatives appearing in the consistent Action~\eqref{eq:actionKin}, as also described in details in Section~\ref{sec:RegIdrawb}.
Applying the four sets of EBC from Eq.~\eqref{courant_3}-\eqref{courant_4} to the solution forms~\eqref{profiles_rein}, it appears that certain constant parameters must be equal to zero  
and thus we obtain the following four possible sets of profiles and KK mass spectrum
equation ($\forall n \geq 0$),
\begin{eqnarray}
1) & (--): \ d_{L}^n(y) = - B^n_{L} \; \sin(m_n \; y), \  (++): \ d_{R}^n(y) = B^n_{L} \; \cos(m_n \; y) \, ; \  \sin(m_n \; L)=0,
\nonumber  \\ 
2) & (++): \ q_{L}^n(y) = B^n_{R} \; \cos(m_n \; y) , \  (--): \ q_{R}^n(y) = B^n_{R} \; \sin(m_n \; y) \, ; \  \sin(m_n \; L)=0,
\nonumber  \\ \label{completeEBCsm0} 
\end{eqnarray}
and,
\begin{eqnarray}
3)  & (-+): \ f_{L}^n(y) = - B^n_{L} \; \sin(m_n \; y), \  (+-): \ f_{R}^n(y) = B^n_{L} \; \cos(m_n \; y) \, ; \  \cos(m_n \; L)=0,
\nonumber  \\ 
4)  & (+-): \ f_{L}^n(y) = B^n_{R} \; \cos(m_n \; y) , \  (-+): \ f_{R}^n(y) = B^n_{R} \; \sin(m_n \; y) \, ; \  \cos(m_n \; L)=0.
\nonumber  \\ \label{completeEBCcusto0} 
\end{eqnarray}
Here, we have used the standard BC notations, {\it i.e.} $-$ or $+$ for example at $y=0$ stands respectively for the Dirichlet or Neumann wave function 
BC: $f_{L/R}^n(0) = 0$ or $\partial_y f_{L/R}^n(0) = 0$. For instance, the symbolic notation $(-+)$ denotes Dirichlet (Neumann) BC at $y=0$ ($y=L$). 
The solutions~\eqref{completeEBCsm0} assigned to the (singlet / doublet component) quark fields give rise to the chiral nature of the SM and to its correct chirality configuration,
as described in Section~\eqref{brTermSec} for Eq.~\eqref{completeEBCsm0}.
The other solutions~\eqref{completeEBCcusto0} lead to KK towers without zero-modes like custodian states [see also the discussion on Eq.~\eqref{completeEBCcusto0} in Section~\eqref{brTermSec}].

Notice that the used BC~\eqref{courant_4} must be injected into the equations~\eqref{ELE_2} issued from the EOM as those are valid for any point of the extra dimension including the boundaries
[see the original Eq.~\eqref{HVP_2}]. This leads to a new set of BC that we call the complete BC. These complete BC are well satisfied by the final solutions~\eqref{completeEBCsm0} and \eqref{completeEBCcusto0}.

The constants $B^n_{L}= \sqrt{2} \, e^{i\alpha^n_L}$ and $B^n_{R}= \sqrt{2} \, e^{i\alpha^n_R} \ (\forall n > 0)$~\footnote{For the solution 1, we find $B^0_{L}= e^{i\alpha^0_L}$ while 
$B^0_{R}= e^{i\alpha^0_R}$ for the solution 2 [{\it c.f.} Eq.~\eqref{completeEBCsm0}].}, where $\alpha^n_{L/R}$ are real angles, are 
fixed by the ortho-normalisation condition~\eqref{normalization_1}. 
The relation $\sin(m_n L)=0$ has the following chosen solutions for the KK mass spectrum, 
\begin{eqnarray}
 \fbox{$m_n=+\frac{n\pi}{L}$}  \  (\forall n \geq 0) \ .
\label{PossSolSpctr} 
\end{eqnarray}
The spectrum $m_n=-\frac{n\pi}{L}$ [$\forall n \geq 0$] is also possible.
Similarly, the relation $\cos(m_n L)=0$ has the possible solutions $m_n=\pm (\frac{\pi}{2L}+\frac{n\pi}{L})$ [$\forall n \geq 0$].

\paragraph{\textit{The current from equations of motion}} $\phantom{0}$ \vspace{0.3cm} \\
Alternatively, as the starting point, one can apply the vanishing conditions~\eqref{courant_2} (EBC) on the same probability currents~\eqref{courant_1} (up to the definition constant $-\alpha^{(\prime)}$) 
satisfying the conservation relations, $\partial_M j_F^M=0$, as derived directly (without applying the Noether theorem) from a 
rewri\-ting~\footnote{Subtracting the Dirac equation to its Hermitian conjugate form, with the relevant 5D field and $\gamma^0$ factors, and using the 5D Dirac matrix rules.} of 
each free 5D Dirac equation~\eqref{ELE_1} in the bulk: 
\begin{equation}
i \Gamma^M \partial_M Q   = 0 \ , \ \ \  
i \Gamma^M \partial_M D   = 0  \ .
\label{ELE_1BB1}
\end{equation}
In order to derive possible NBC, one has now to consider the action. 
The free bulk fermion action can be rewritten, after an integration by part in the last two terms of Eq.~\eqref{L_3}, as
\begin{eqnarray}
{S}_{\Psi} & = & \int d^4x \, dy
\sum_{F = Q, D} \left\{ i F^\dagger_R \sigma^\mu \partial_\mu F_R + i F^\dagger_L \bar{\sigma}^\mu \partial_\mu F_L 
+ F^\dagger_R \partial_4 F_L - F^\dagger_L \partial_4 F_R  \right.
\nonumber \\ &  & \ \ \ \ + \left. \dfrac{1}{2} \left[ \delta(y-L) - \delta(y) \right] \left( F^\dagger_L F_R - F^\dagger_R F_L \right)  \right\} \, .
\label{modified_action_1}
\end{eqnarray}

Injecting directly the EBC~\eqref{courant_2} into the Lagrangian would cancel out the boundary terms of Eq.~\eqref{modified_action_1} and in turn spoil the necessary hermiticity of the action [being explicit through 
Eq.~\eqref{eq:actionKin}-\eqref{L_3}]. This feature reflects the fact that the action and the current condition~\eqref{courant_2} are distinct ingredients defining the model. The proper method goes as follows;
the current condition~\eqref{courant_2} constitutes the EBC which will have to be combined with the action minimisation relations.

So let us apply the least action principle to the Action~\eqref{modified_action_1}. From the known EBC~\eqref{courant_2}, leading to the conditions~\eqref{courant_3}, we deduce that the field $\left. F_L \right |_{0}$ or $\left. F_R \right |_{0}$ 
is fixed to zero, as well as the field $\left. F_L \right |_{L}$ or $\left. F_R \right |_{L}$, so that their functional variation vanishes [as in Eq.~\eqref{delta-BC}] and in turn the global variation of the action part in the
second line of Eq.~\eqref{modified_action_1} cancels out. Another way to find out this cancellation is to combine the functional variation of the relation~\eqref{courant_2} with the variation of the terms in the second line 
of Eq.~\eqref{modified_action_1}. Therefore, the action minimisation only leads to the same bulk EOM as in Eq.~\eqref{ELE_1}. These bulk equations induce the profile equations~\eqref{ELE_2} whose solutions 
have the forms given by Eq.~\eqref{profiles_rein}, which are taken once more continuous at the boundaries. As above, the EBC~\eqref{courant_2}-\eqref{courant_3}-\eqref{courant_4} applied to these solutions give 
rise to the final profiles~\eqref{completeEBCsm0}-\eqref{completeEBCcusto0} and in turn to the mass spectrum discussed via Eq.~\eqref{PossSolSpctr}.

\subsubsection{Introducing the bilinear brane terms [NBC]}
\label{sec:freeNBC}

As announced at the end of Section~\ref{no_Yukawa}, 
an alternative method~\footnote{One could simultaneously impose the EBC~\eqref{courant_2} above and add the BBT to the action, but this method would contain some redundancy.} with respect to previous subsection
for finding out the same consistent physical solutions, for the mass spectrum and the profiles, is to add the BBT~\eqref{eq:actionBound} to the kinetic terms~\eqref{L_3} so that the initial free fermionic action becomes,
\begin{eqnarray}
S^{\rm{free}}_{\rm{5D}} = S_{\Psi} + S_B \ . \label{eq:actionFreeBBT}
\end{eqnarray}
Let us apply the least action principle using this action as the starting point. 
The action based on Eq.~\eqref{L_3}, rewritten as the Action~\eqref{modified_action_1} and added to the BBT piece~\eqref{eq:actionBound}, reads as, after combining the boundary terms,
\begin{eqnarray}
S^{\rm{free}}_{\rm{5D}} =   \int d^4x \, dy
\sum_{F = Q, D} \left\{ i F^\dagger_R \sigma^\mu \partial_\mu F_R + i F^\dagger_L \bar{\sigma}^\mu \partial_\mu F_L
+ F^\dagger_R \partial_4 F_L - F^\dagger_L \partial_4 F_R \right\} 
\nonumber  \\ 
+ \   \int d^4x \, dy \  [ \delta(y-L) - \delta(y) ] \ \left( Q^\dagger_L Q_R - D^\dagger_R D_L \right) \ .
\nonumber
\end{eqnarray}
Without loss of generality, the stationary action condition can be split into these two conditions with respect to the two field variations respectively,
\begin{eqnarray}
0 = \delta_{F^\dagger_L} S^{\rm{free}}_{\rm{5D}} & = & \displaystyle{ \int d^4x \; \int_0^{L} dy \;  
\left[ \delta F^\dagger_L \; i\bar{\sigma}^\mu \partial_\mu F_L - \delta F^\dagger_L \; \partial_4 F_R  \right] } 
\nonumber \\
& & \displaystyle{ + \ \int d^4x \; C^L_F \, \left[  \left . \left( \delta F^\dagger_L \; F_R \right) \right |_{L} - \left . \left( \delta F^\dagger_L \; F_R \right) \right |_{0} \; \right] } \ ,
\ \label{HVP_3a} \\
0 = \delta_{F^\dagger_R} S^{\rm{free}}_{\rm{5D}} & = & \displaystyle{ \int d^4x \; \int_0^{L} dy \;  
\left[ \delta F^\dagger_R \; i\sigma^\mu \partial_\mu F_R + \delta F^\dagger_R \; \partial_4 F_L  \right] } 
\nonumber \\
& & \displaystyle{ + \ \int d^4x \; C^R_F \,  \left[ - \left . \left( \delta F^\dagger_R \; F_L \right) \right |_{L} + \left . \left( \delta F^\dagger_R \; F_L \right) \right |_{0} \; \right] } \ ,
\ \label{HVP_3b}
\end{eqnarray}
where $C^L_D=0$, $C^R_D=1$, $C^L_Q=1$, $C^R_Q=0$, for each field $D$, $Q$. For generic field variations $\delta F^\dagger_{L/R}$ and $\left . \delta F^\dagger_{L/R} \right |_{0, L}$, 
the sum of the first two terms, both in Eq.~\eqref{HVP_3a} and~\eqref{HVP_3b}, must vanish separately, leading to the same equations as the EOM~\eqref{ELE_1}~\footnote{We 
obtain the Hermitian conjugate EOM and NBC by integrating by part the bulk piece of  
the relation $\delta_{F_{L,R}} S^{\rm{free}}_{\rm{5D}} = 0$ (non-vanishing boundary terms appear due to the integration over the extra dimension) in order to get rid of the field factors $\partial_M \delta F_{L,R}$.} 
and in turn via Eq.~\eqref{Dirac_1} to the profile equations~\eqref{ELE_2} with solutions~\eqref{profiles_0}. 
The general solutions~\eqref{profiles_0}, once injected into the initial equations~\eqref{ELE_2}, take the specific forms~\eqref{profiles_rein}.
We are thus left with the NBC: 
\begin{equation}
C^L_F \left . \left( \delta F^\dagger_L \; F_R \right) \right |_{L}  = 0 , \ 
C^L_F     \left . \left( \delta F^\dagger_L \; F_R \right) \right |_{0}    = 0 , \ 
C^R_F     \left . \left( \delta F^\dagger_R \; F_L \right) \right |_{L}   = 0 \phantom{0} \text{and} \phantom{0}
C^R_F   \left . \left( \delta F^\dagger_R \; F_L \right) \right |_{0}  = 0 \  .
\label{HVP_4n}
\end{equation}
Then using the appropriate constants $C^{L,R}_Q, C^{L,R}_D$ above for each field and generic variations $\left . \delta F^{(\dagger)}_{L,R}  \right |_{0,L}  \neq 0$, 
it appears clearly that those BC belong to the set of BC~\eqref{courant_3}-\eqref{courant_4} whose application on the solution forms~\eqref{profiles_rein} leads to
the two respective sets of profiles and KK mass spectrum~\eqref{completeEBCsm0}, as already derived.
The structure of the profile solutions~\eqref{completeEBCsm0} corresponds to the chiral nature and configuration of the SM as already explained in Section~\ref{brTermSec}.

For completeness, starting instead from the BBT~\eqref{eq:actionBoundCUSTO} in the action for a given field $F$,
\begin{eqnarray}
S^{\prime\rm{free}}_{\rm{5D}} = S_{\Psi} (F) + S'_B \ , \label{eq:actionFreeBBTcusto}
\end{eqnarray}
the combination of the boundary terms leads to,
\begin{eqnarray}
S^{\prime\rm{free}}_{\rm{5D}}& =  & \int d^4x \, dy \
 \left\{ i F^\dagger_R \sigma^\mu \partial_\mu F_R + i F^\dagger_L \bar{\sigma}^\mu \partial_\mu F_L
+ F^\dagger_R \partial_4 F_L - F^\dagger_L \partial_4 F_R \right . 
\nonumber  \\ \nonumber 
& + &   \left .  \left [ \delta(y-L) \frac{\sigma+1}{2} + \delta(y)  \frac{\sigma-1}{2}  \right ] \ F^\dagger_L F_R \ 
+ \   \left  [ \delta(y-L)  \frac{\sigma-1}{2} + \delta(y) \frac{\sigma+1}{2}  \right ] \ F^\dagger_R F_L  \right\}  \ .
\label{L_4custo}
\end{eqnarray}
The stationary action condition can be split into the two following conditions, 
\begin{eqnarray}
0 = \delta_{F^\dagger_L} S^{\prime\rm{free}}_{\rm{5D}} & = & \displaystyle{ \int d^4x \; \int_0^{L} dy \;  
\left[ \delta F^\dagger_L \; i\bar{\sigma}^\mu \partial_\mu F_L - \delta F^\dagger_L \; \partial_4 F_R  \right] } 
\nonumber \\
& & \displaystyle{ + \ \int d^4x \;  \left[  \left .  \frac{\sigma+1}{2} \left( \delta F^\dagger_L \; F_R \right) \right |_{L} + \left .  \frac{\sigma-1}{2}  \left( \delta F^\dagger_L \; F_R \right) \right |_{0} \; \right] } \ ,
\ \label{HVP_3acusto} \\
0 = \delta_{F^\dagger_R} S^{\prime\rm{free}}_{\rm{5D}} & = & \displaystyle{ \int d^4x \; \int_0^{L} dy \;  
\left[ \delta F^\dagger_R \; i\sigma^\mu \partial_\mu F_R + \delta F^\dagger_R \; \partial_4 F_L  \right] } 
\nonumber \\
& & \displaystyle{ + \ \int d^4x \;  \left[ \left .  \frac{\sigma-1}{2} \left( \delta F^\dagger_R \; F_L \right) \right |_{L} + \left .  \frac{\sigma+1}{2}  \left( \delta F^\dagger_R \; F_L \right) \right |_{0} \; \right] } \ .
\ \label{HVP_3bcusto}
\end{eqnarray}
Once more, the sum of the first two terms in Eq.~\eqref{HVP_3acusto} and~\eqref{HVP_3bcusto}, respectively, must vanish, leading to the same profile equations as the ones deduced from Eq.~\eqref{HVP_3a}-\eqref{HVP_3b} 
and hence to the identical bulk solution forms~\eqref{profiles_rein}. Nevertheless, we are now left with the new NBC: 
\begin{equation}
\frac{\sigma+1}{2}  \left . \left( \delta F^\dagger_L \; F_R \right) \right |_{L}  =   
\frac{\sigma-1}{2}  \left . \left( \delta F^\dagger_L \; F_R \right) \right |_{0}    =   
\frac{\sigma-1}{2} \left . \left( \delta F^\dagger_R \; F_L \right) \right |_{L}   =  
\frac{\sigma+1}{2}  \left . \left( \delta F^\dagger_R \; F_L \right) \right |_{0}  = 0 \  .
\label{HVP_4ncusto}
\end{equation}
Then for generic variations $\left . \delta F^{(\dagger)}_{L,R}  \right |_{0,L}  \neq 0$, 
it is clear that those BC belong to the set of BC~\eqref{courant_3}-\eqref{courant_4} whose application on the solution forms~\eqref{profiles_rein} leads, for $\sigma =+1$, to
the set 3 of profiles and KK mass spectrum in Eq.~\eqref{completeEBCcusto0}, and, for $\sigma =-1$, to the set 4 of solutions in Eq.~\eqref{completeEBCcusto0}, as already derived.
The control of the BBT sign factor $\sigma$, in Eq.~\eqref{eq:actionBoundCUSTO}, on the final solution structure appears here clearly.
The profile solutions~\eqref{completeEBCcusto0} have a custodian chiral structure as already described in Section~\ref{brTermSec}.

\paragraph{\textit{The current from action variations}} $\phantom{0}$ \vspace{0.3cm} \\
In the presence of the BBT~\eqref{eq:actionBound} or \eqref{eq:actionBoundCUSTO} [invariant under the transformations~\eqref{symetrie_1}], 
as demonstrated in the beginning of Appendix~\ref{Noether_THM}, the application of the Noether theorem based on the bulk EOM~\eqref{ELE_1} -- derived from the 
variation of the Action~\eqref{eq:actionKin}-\eqref{L_3} invariant under the global $U(1)_F$ transformations~\eqref{symetrie_1} -- leads to the same probability currents~\eqref{currentGEN} defined separately 
for the bulk fermions represented by the 5D fields $F=Q,D$, as in Eq.~\eqref{courant_1}.
Now the NBC~\eqref{HVP_4n} or \eqref{HVP_4ncusto} induced by the BBT, as both satisfying the BC~\eqref{courant_3}, lead to 4 conditions of vanishing probability currents of the exact form~\eqref{courant_2}.
In other words, the presence of the BBT guarantees (without imposing any condition) the vanishing of the currents at both boundaries for each independent fermion species.
These BBT-induced conditions contribute to the consistent and complete definition of the geometrical field configuration for the considered model with a finite extra spatial interval in which fermionic matter is stuck.

\paragraph{\textit{The current from equations of motion}} $\phantom{0}$ \vspace{0.3cm} \\
Alternatively, we can derive directly (without the Noether theorem) the conservation relations, $\partial_M j_F^M=0$, for the probability currents~\eqref{courant_1} from a 
rewriting~\footnote{Subtracting the Dirac equation to its Hermitian conjugate form.} of each free 5D Dirac equation~\eqref{ELE_1} in the bulk as in Eq.~\eqref{ELE_1BB1}. 
The BBT~\eqref{eq:actionBound} or \eqref{eq:actionBoundCUSTO} affect only the NBC derived from variation of the Action~\eqref{eq:actionKin}-\eqref{L_3}.
The NBC~\eqref{HVP_4n} or \eqref{HVP_4ncusto}, induced by the BBT, both respect the BC~\eqref{courant_3}, and hence lead to the 4 conditions of vanishing currents~\eqref{courant_2}.
Therefore, as above, the BBT guarantee the vanishing of the currents at both boundaries for each fermion species.
Those conditions allow a consistent and complete definition of the geometrical field set-up for the model with matter on an interval. 

\vspace{0.3cm}

Let us close this part by remarking that one could as well combine the two approaches to define the model: add a BBT only on an interval boundary for a given 5D field (as in this Section~\ref{sec:freeNBC})
and apply the vanishing current condition only on the other boundary (as in Section~\ref{sec:freeEBC}).

\subsubsection{Presence of Yukawa couplings}
\label{with_Yukawa}

In this section, we only describe the two steps of a first 
method~\cite{Huber:2000fh, delAguila:2000kb, Hewett:2002fe, Huber:2003tu, Agashe:2006iy, Agashe:2006wa, Djouadi:2007fm, Goertz:2008vr, Bouchart:2009vq, Barcelo:2014kha}, 
that will turn out to be the correct method, 
for including the effects of the Yukawa terms~\eqref{L_5} on the final fermion spectrum. 
First, the free profiles and free spectrum are calculated within a strict 5D approach whose correct treatment was exposed in details in Sections~\ref{sec:freeEBC} and ~\ref{sec:freeNBC}.
Secondly, one can write a mass matrix for the 4D fermion fields involving the pure KK masses [the free spectrum of first step] as well as the masses induced by the Higgs 
VEV in the Yukawa terms~\eqref{L_5} [with free profiles of first step] which mix together the KK modes. The bi-diagonalisation of this matrix gives rise to an infinite set of eigenvalues 
constituting the physical masses, as will be presented in Section~\ref{Mass_matrix_diagonalisation}.

\subsection{The 4D fermion mass matrix}
\label{Mass_matrix_diagonalisation}

We focus on the fermion terms of the 5D Action~\eqref{eq:actionTot} in order to work out the mass spectrum: in particular on $S_{\Psi}$, $S_X$ and possibly $S_B$ (without direct effect on the mass 
matrix~\eqref{M} as explained in Section~\ref{brTermSec}) if no EBC are applied. 
Those terms lead -- after insertion of the KK decomposition~\eqref{KK_1}, use of free Eq.~\eqref{ELE_2},
ortho-normalisation condition~\eqref{normalization_1} and integration over the fifth dimension -- to the canonical kinetic terms for the 4D fermion fields as well as to  
the following fermionic 4D effective mass terms in the Lagrangian density (and to independent 4D effective Higgs-fermion couplings not discussed here),
\begin{equation}
- \chi^\dagger_L \; \mathcal{M} \; \chi_R + \text{H.c.} \nonumber
\end{equation}
in the combined basis for the left and right-handed (transposed) 4D fields:
\begin{equation}
\left\{
\begin{array}{r c l}
\chi^t_L &=& \left( Q_L^{0t}, Q_L^{1t}, D_L^{1t}, Q_L^{2t}, D_L^{2t}, \cdots \right) \\ \\
\chi^t_R &=& \left( D_R^{0t}, Q_R^{1t}, D_R^{1t}, Q_R^{2t}, D_R^{2t}, \cdots \right)    
\label{basis}
\end{array}
\right.  \ \ \ .
\end{equation}
Notice that there exists only one chirality for the zero-modes as explained below Eq.~\eqref{eq:actionBound}. 
The infinite mass matrix reads as
\begin{equation}
\mathcal{M} =
\begin{pmatrix}
\alpha_{00} & 0 & \alpha_{01} & 0 & \alpha_{02} & \cdots \\
\alpha_{10} & m_{1} & \alpha_{11} & 0 & \alpha_{12} & \cdots \\
0 & \beta_{11} & m_{1} & \beta_{12} & 0 & \cdots \\
\alpha_{20} & 0 & \alpha_{21} & m_{2} & \alpha_{22} & \cdots \\
0 & \beta_{21} & 0 & \beta_{22} & m_{2} & \cdots \\
\vdots & \vdots & \vdots & \vdots & \vdots & \ddots \\
\end{pmatrix}
\label{M}
\end{equation}
where $m_n$ is the free spectrum~\eqref{PossSolSpctr} 
and the free wave function overlaps with the Higgs-brane are defined by (for real wave functions)
\begin{equation}
\left\{
\begin{array}{l}
\forall (i, j) \in \mathbb{N}^2, \; \alpha_{ij} = \dfrac{X}{L} \; q_L^{i}(L) \; d_R^j(L) \\ 
\forall (i, j) \in \mathbb{N}^{\star 2}, \; \beta_{ij} = \dfrac{X'}{L} \; d_L^{i}(L) \; q_R^j(L)
\end{array}
\right.   \ \ \ .  
\label{coeff}
\end{equation}
As the profiles are the free ones [profiles and KK mass spectrum solutions~\eqref{completeEBCsm0} 
with SM chiral structure], only the $\alpha_{ij}$ coefficients do not vanish.

The physical fermion mass spectrum is obtained by bi-diagonalising the mass matrix~\eqref{M}.
This method is called the perturbation method in the sense that truncating the mass matrix at a given KK level corresponds to keeping only the dominant contributions
to the lightest mass eigenvalue being the measured fermion mass (higher KK modes tend to mix less to the zero-mode due to larger mass differences).

Extracting the mass spectrum equation from the characteristic equation for the Hermi\-tian-squared mass matrix~\eqref{M}, in the case of infinite KK towers, is not trivial. 
This useful exercise was addressed analytically in Ref.~\cite{Barcelo:2014kha} for the present toy model but with a 5D Yukawa coupling constant (and in turn a $X$ quantity) 
taken real. The resulting exact equation -- without any approximation -- was found to be:
\begin{equation}
\tan^2(\sqrt{\vert M_n \vert^2} \; L) = X^2 \ \Leftrightarrow \ \tan(\sqrt{\vert M_n \vert^2} \; L) = \pm X  \ \ (\forall n \geq 0) \  .
\label{4D_spectrum}
\end{equation}
Let us present here the absolute values of the solutions (physical masses) of this equation:
\begin{equation}
\vert M_n \vert = \left \vert \dfrac{\arctan(\pm X) +(-1)^n \, \tilde n(n) \, \pi}{L} \right \vert \ \ (\forall n \geq 0) 
\label{4D_spectrumSOL}
\end{equation}
where the function $\tilde n(n)$ is defined by
\begin{equation}
\left\{
\begin{array}{r c l}
\tilde n(n) &=& \frac{n}{2} \ {\rm for} \ n \ {\rm even}
\\ \vspace{-0.2cm} \\
\tilde n(n) &=& \frac{n+1}{2} \ {\rm for} \ n \ {\rm odd}
\end{array}
\right. 
\label{ntilde}
\end{equation}
so that the positive integer $n$ labelling the mass eigenvalues remains as well the label of the associated [as in the free case~\eqref{Dirac_1}] 4D mass eigenstates $\psi^n(x_\mu)$ [like in Eq.~\eqref{KK_1}].
Besides, globally speaking, for the whole set of $n$-levels, the absolute value of the fermion mass has a common generic expression for the two different 
classes ($\pm X$)~\footnote{The other solution, consisting alternatively of Eq.~\eqref{4D-prof-BC_Y_4} for some $n$-levels and Eq.~\eqref{4D-prof-BC_Y_4bis} for other $n$-levels, has to be ruled out since 
the complete and consistent infinite mass spectrum solution is fixed in a unique model hypothesis selected among the two 
given mathematical solutions~\eqref{4D-prof-BC_Y_4} or~\eqref{4D-prof-BC_Y_4bis}, as Eq.~\eqref{4D_spectrum} determines the quantity $\sqrt{\vert M_n \vert^2} L$ modulo $\pi$.} of solutions: 
\begin{eqnarray}
\tan(\sqrt{\vert M_n \vert^2} \; L) = + X & \Rightarrow & \vert M_n \vert = \fbox{$ \left \vert \dfrac{\arctan(X) + (-1)^n \, \tilde n(n) \, \pi}{L} \right \vert $} \ , \label{4D-prof-BC_Y_4} \\
{\rm \ or, \ \ } \tan(\sqrt{\vert M_n \vert^2} \; L) = - X & \Rightarrow & \vert M_n \vert =  \dfrac{1}{L} \left \vert  -\arctan(X) + (-1)^n \, \tilde n(n) \, \pi \right \vert \label{4D-prof-BC_Y_4bis} \\ 
&& \ \ \ \ \ \ \ = \dfrac{1}{L} \left \vert  \arctan(X) + (-1)^{n+1} \, \tilde n(n) \, \pi  \right \vert \nonumber \\ 
&& \ \ \ \ \ \ \ \equiv \fbox{$ \left \vert \dfrac{\arctan(X) + (-1)^n \, \tilde n(n) \, \pi}{L} \right \vert $} \ .
\nonumber
\end{eqnarray}
The last equality is justified for the whole spectrum by the fact that, for two consecutive $n$ values 
(for one odd $n$ and the following even $n$, with $n>0$~\footnote{The justification is obvious (no sign effect) for the special $n$-even case $\tilde n(0)=0$.}), 
$(-1)^n$ and $(-1)^{n+1}$ span the same two values $\pm 1$ while $\tilde n(n)$ keeps the same value. 
Hence the two classes of solutions in Eq.~\eqref{4D-prof-BC_Y_4} and \eqref{4D-prof-BC_Y_4bis} may only differ by some different mass signs (remaining unfixed in the solutions for absolute values).

To check that the counting of states is correct, we observe that, in the realistic case $\vert X \vert \ll 1$ 
(typically small SM masses compared to the KK scale), two consecutive absolute masses $\vert M_n \vert$ 
(for one odd $n$ and the following even $n$, with $n>0$) of Eq.~\eqref{4D_spectrumSOL} are equal at leading order to the corresponding [unique $\tilde n$ value] absolute mass 
$\vert \pm \tilde n \, \pi / L \vert$ as in the free spectrum~\eqref{PossSolSpctr}. Hence, in the vanishing mixing limit [see matrix~\eqref{M}], the two associated consecutive mass 
eigenstates $\psi^n(x_\mu)$ tend well to the two free 4D field components $Q^{\tilde n}(x_\mu)$ and $D^{\tilde n}(x_\mu)$ [of Eq.~\eqref{Dirac_1}].

\section{5D Treatment: the regularisation doom}
\label{Usual_treatment}

In this part, we work out the fermion mass spectrum in the defined model with the 5D Action~\eqref{eq:actionTot} using 
the alternative 5D approach based on the brane-Higgs 
regularisation~\cite{Csaki:2005vy,Grojean:2007zz,Barcelo:2014kha,Csaki:2003sh,Casagrande:2008hr,Azatov:2009na,Casagrande:2010si} 
and we point out non rigorous patterns of this method.

\subsection{Mixed Kaluza-Klein decomposition}
\label{5D_approach}

As we have just seen in Eq.~\eqref{basis}-\eqref{M}, after EW symmetry breaking, the infinite $Q^n_L$ and $D^n_L$ field towers mix together (as well as the $Q^n_R$ and $D^n_R$) 
to form 4D fields $\psi^n_L$ (and $\psi^n_R$) representing mass eigenstates. In order to take into account this mixing within the 5D approach, these common 4D fields $\psi^n_{L}$
are defined instead of the $Q^n_L$ and $D^n_L$ fields (and similarly for the right-handed fields) in the whole KK decomposition, then called a \textit{mixed} KK 
decomposition [instead of the \textit{free} one in Eq.~\eqref{KK_1}]~\cite{Azatov:2009na}, as follows, 
\begin{equation}
\left\{
\begin{array}{r c l}
Q_L \left( x_\mu, y \right) &=& \dfrac{1}{\sqrt{L}} \displaystyle{ \sum^{+\infty}_{n=0} q^n_L(y) \; \psi^n_L \left( x_\mu \right) }
\\ \vspace{-0.3cm} \\
Q_R \left( x_\mu, y \right) &=& \dfrac{1}{\sqrt{L}} \displaystyle{ \sum^{+\infty}_{n=0} q^n_R(y) \; \psi^n_R \left( x_\mu \right) }
\\ \vspace{-0.3cm} \\
D_L \left( x_\mu, y \right) &=& \dfrac{1}{\sqrt{L}} \displaystyle{ \sum^{+\infty}_{n=0} d^n_L(y) \; \psi^n_L \left( x_\mu \right) }
\\ \vspace{-0.3cm} \\
D_R \left( x_\mu, y \right) &=& \dfrac{1}{\sqrt{L}} \displaystyle{ \sum^{+\infty}_{n=0} d^n_R(y) \; \psi^n_R \left( x_\mu \right) }
\end{array} 
\right. 
\label{KK_2}
\end{equation}
The 4D fields $\psi^n_{L/R}$ ($\forall n$) must satisfy the Weyl equations
\begin{equation}
\left\{
\begin{array}{r c l}
i \bar{\sigma}^\mu \partial_\mu \psi^n_L \left( x_\mu \right) &=& M_n \; \psi_R^n \left( x_\mu \right)
\\ \vspace{-0.2cm} \\
i \sigma^\mu \partial_\mu \psi^n_R \left( x_\mu \right) &=& M_n \; \psi_L^n \left( x_\mu \right)
\end{array}
\right.
\label{Dirac_2}
\end{equation}
where the spectrum $M_n$ includes the mass contribution whose origin is the Yukawa couplings~\eqref{L_5}. Note that in contrast with the free case, 
there is a unique mass spectrum $M_n$ for a unique 4D field tower $\psi^n_{L/R} (x_\mu)$.
In order to guarantee the existence of diagonal and canonical kinetic terms for those 4D fields $\psi^n_{L/R}$, the associated new profiles must now obey the two following 
ortho-normalisation conditions,
\begin{equation}
\dfrac{1}{L} \int_0^{L} dy \; \left[ q^{n\star}_{\cal C}(y) \; q^m_{\cal C}(y) + d^{n\star}_{\cal C}(y) \; d^m_{\cal C}(y) \right] = \delta_{nm} \ , \forall n \ , \forall m \ ,
\label{normalization_2}
\end{equation}
for a chirality index ${\cal C} \equiv L$ or $R$. These two conditions are different from the four ones of Eq.~\eqref{normalization_1} due to the new mixed KK decomposition.

\subsection{Inconsistencies of the Higgs shift procedure}
\label{RegInterval}

Here we highlight the formal problems of the 5D process of shifting the brane-Higgs field~\cite{Barcelo:2014kha,Csaki:2003sh,Casagrande:2010si} 
to get the fermion mass tower. Once more the considered fermion terms of the 5D 
Action~\eqref{eq:actionTot} are $S_{\Psi}$ and $S_X$ (without $S_B$ which was missed in the relevant literature and that will be taken into account 
in Section~\ref{Yukawa_terms_as boundary_conditions}). The variations of the studied action lead to the same free BC [second line of Eq.~\eqref{HVP_2}] but to the following bulk EOM 
including the Yukawa coupling constants [instead of the free ones in Eq.~\eqref{ELE_1}]~\footnote{In the subsections on Higgs regularisations, we use the same Lagrangians as in the present paper but 
the results and conventions from Ref.~\cite{Barcelo:2014kha}.}, 
\begin{equation}
\left\{
\begin{array}{r c l}
i \bar{\sigma}^\mu \partial_\mu Q_L + \partial_4 Q_R + \delta(y-L) \; X \; D_R &=& 0\ ,
\\ \vspace{-0.2cm} \\
i \sigma^\mu \partial_\mu Q_R - \partial_4 Q_L + \delta(y-L) \; X' \; D_L &=& 0\ ,
\\ \vspace{-0.2cm} \\
i \bar{\sigma}^\mu \partial_\mu D_L + \partial_4 D_R + \delta(y-L) \; X' \; Q_R &=& 0\ ,
\\ \vspace{-0.2cm} \\
i \sigma^\mu \partial_\mu D_R - \partial_4 D_L + \delta(y-L) \; X \; Q_L &=& 0 \ .
\end{array}
\right.
\label{ELE_4}
\end{equation}
Indeed, in view of regularising the brane-Higgs field, the Yukawa interactions must be included in the bulk EOM~\cite{Barcelo:2014kha} -- as done in the literature.
Inserting the mixed KK decomposition~\eqref{KK_2} in these 5D field EOM~\eqref{ELE_4} allows to factorise out the 4D fields, obeying the 4D Dirac equations~\eqref{Dirac_2}, 
and obtain the profile equations for each excited mode [instead of the free ones in Eq.~\eqref{ELE_2}]:
\begin{equation}
\forall n\ , \ 
\left\{
\begin{array}{r c l}
\partial_4 q^n_R(y) + M_n \; q^n_L(y) &=& - \ \delta(y-L) \; X \; d^n_R(y) \ ,
\\ \vspace{-0.2cm} \\
\partial_4 q^n_L(y) - M_n \; q^n_R(y) &=& \delta(y-L) \; X' \; d^n_L(y) \ ,
\\ \vspace{-0.2cm} \\
\partial_4 d^n_R(y) + M_n \; d^n_L(y) &=& - \ \delta(y-L) \; X' \; q^n_R(y) \ ,
\\ \vspace{-0.2cm} \\
\partial_4 d^n_L(y) - M_n \; d^n_R(y) &=& \delta(y-L) \; X \; q^n_L(y) \ .
\end{array}
\right.
\label{ELE_5}
\end{equation}
Here we underline a first mathematical issue of this usual approach: 
introducing $\delta(y-L)$ Dirac peaks~\footnote{Strictly speaking, a Dirac peak is a distribution although its historical name is ``Dirac delta function''.} 
in these profile equations leads to relations between distributions~\footnote{Also called ``generalised functions'' in mathematical analysis.} 
and functions which are thus not mathematically consistent~\cite{Schwartz1,Schwartz2}.

The apparent ``ambiguity'' noticed in the literature (context of a warped extra dimension) was that the Yukawa terms in Eq.~\eqref{ELE_5} are present only at the $y=L$ boundary
and might thus affect the fermion BC. In order to avoid this question of a potential problem (like a field vagueness), a regularisation of the brane-Higgs coupling was suggested 
forcing to keep the free fermion BC in the presence of Yukawa interactions.

\subsubsection{Regularisation I drawbacks}
\label{sec:RegIdrawb}

In the first type of regularisation applied in the literature~\cite{Casagrande:2008hr,Csaki:2003sh,Barcelo:2014kha}, called Regula\-risation~I, the BC are considered at the first level of the 
procedure to be injected in Eq.~\eqref{ELE_5}~\cite{Barcelo:2014kha}. The free BC impose $d^n_L(L)=q^n_R(L)=0$ [see respectively the first and fourth solutions in  
Eq.~\eqref{completeEBCsm0}] so that Eq.~\eqref{ELE_5} is supposed to become
\begin{equation}
\forall n\ , \ 
\left\{
\begin{array}{r c l}
\partial_4 q^n_R(y) + M_n \; q^n_L(y) &=& - \ \delta(y-L) \; X \; d^n_R(y) \ ,
\\ \vspace{-0.2cm} \\
\partial_4 q^n_L(y) - M_n \; q^n_R(y) &=& 0 \ ,
\\ \vspace{-0.2cm} \\
\partial_4 d^n_R(y) + M_n \; d^n_L(y) &=& 0 \ ,
\\ \vspace{-0.2cm} \\
\partial_4 d^n_L(y) - M_n \; d^n_R(y) &=& \delta(y-L) \; X \; q^n_L(y) \ .
\end{array}
\right.
\label{ELE_5bis}
\end{equation}
At this level, we point out a second lack of strictness in the standard treatment; the two vanishing right-hand sides of Eq.~\eqref{ELE_5bis} originate from the assumption 
that $0 \times \delta(0)=0$ whereas the quantity 
$0 \times \delta(0)$ is rigorously undefined~\footnote{This quantity corresponds also to an undefined product, namely $0 \times \infty$, within the 
original simplified description~\cite{Dirac:1927we} still used in physics textbooks (together with normalisation conditions):
\begin{equation}
\delta(y-L) \equiv
\left\{
\begin{array}{c c c}
0 &$\text{if}$& y \neq L \\
\infty &$\text{if}$& y = L
\end{array}
\right. \ .
%\label{def_delta_f}
\nonumber
\end{equation}
}
%GMF  
which should forbid to continue this standard method~\footnote{Such $\delta(0)$ divergences are automatically regulated -- by the exchange of infinite towers of KK scalar modes -- 
for a brane-Higgs coupled to bulk scalar fields within a minimal supersymmetric scenario~\cite{Bouchart:2011va}.}. 
In the next step of this method, the usual mathematical trick is to shift the brane-Higgs coupling from the brane at $y=L$  
(TeV-brane in a warped framework) by an amount $\epsilon$:
\begin{equation}
\forall n\ , \ 
\left\{
\begin{array}{r c l}
\partial_4 q^n_R(y) + M_n \; q^n_L(y) &=& - \ \delta(y-[L-\epsilon]) \; X \; d^n_R(y) \ ,
\\ \vspace{-0.2cm} \\
\partial_4 q^n_L(y) - M_n \; q^n_R(y) &=& 0 \ ,
\\ \vspace{-0.2cm} \\
\partial_4 d^n_R(y) + M_n \; d^n_L(y) &=& 0 \ ,
\\ \vspace{-0.2cm} \\
\partial_4 d^n_L(y) - M_n \; d^n_R(y) &=& \delta(y-[L-\epsilon]) \; X \; q^n_L(y) \ .
\end{array}
\right.
\label{ELE_5ter}
\end{equation}
Then the integration of the four relations of Eq.~\eqref{ELE_5ter} over an infinitesimal range, tending to zero and centered at $y=L-\epsilon$, leads 
to~\footnote{The integration of Eq.~\eqref{ELE_5ter} could also be performed over the interval $[L-\epsilon,L]$; this variant of the calculation,
suggested in an Appendix of Ref.~\cite{Csaki:2003sh}, represents in fact an equivalent regularisation process leading to the same physical results and 
with identical mathematical inconsistencies.}
\begin{equation}
\forall n\ , \ 
\left\{
\begin{array}{r c l}
q^n_R([L-\epsilon]^+)-q^n_R([L-\epsilon]^-) &=& - \; X \; d^n_R(L-\epsilon) \ ,
\\ \vspace{-0.2cm} \\
q^n_L([L-\epsilon]^+)-q^n_L([L-\epsilon]^-)   &=& 0 \ ,
\\ \vspace{-0.2cm} \\
d^n_R([L-\epsilon]^+)-d^n_R([L-\epsilon]^-)   &=& 0 \ ,
\\ \vspace{-0.2cm} \\
d^n_L([L-\epsilon]^+)-d^n_L([L-\epsilon]^-) &=& X \; q^n_L(L-\epsilon) \ .
\end{array}
\right.
\label{ELE_5integ}
\end{equation}
Another inconsistency arising here in the regularisation process is the following one. The first and fourth relations in Eq.~\eqref{ELE_5integ} show that the 
wave functions $q^n_R(y)$ and $d^n_L(y)$ possess a discontinuity at $y=L-\epsilon$. Hence the functions $\partial_4 q^n_R(y)$ and $\partial_4 d^n_L(y)$
are not defined at $y=L-\epsilon$. Two of the integrations performed on Eq.~\eqref{ELE_5ter} to get Eq.~\eqref{ELE_5integ} are thus not well defined. The  
fundamental theorem of analysis~\footnote{Let $(a, b) \in \mathbb{R}^2$ and $g$ be a continuous function on $[a, b]$, then $g$ admits continuous primitives on $[a, b]$. 
Let $G$ be one of them, then one has: $\int_a^b dy \; g(y) = G(b)-G(a)$.}~\cite{Burk} cannot be applied for functions undefined on the whole interval of integration.
Let us express this problem in other terms; the functions $\partial_4 q^n_R(y)$ and $\partial_4 d^n_L(y)$ being not defined at $y=L$ 
(in the limit $\epsilon \to 0$), the last two terms of the starting 5D Action~\eqref{L_3} -- defined along the interval $\mathcal{C}^1 \equiv  [0, L]$ -- are not well 
defined~\footnote{From the current point of view, the conservation condition~\eqref{currentGENYuk} -- involving in parti\-cular the 5D 
probability current component~\eqref{courant_2Y} -- cannot be properly written at any point along the fifth dimension since $q^n_R(y)$ and $d^n_L(y)$ 
have discontinuities at $y=L$ so that derivatives in $\partial_4 j^4$ are not well defined there.}.
Another definition problem appears in this regularisation; 
the Action~\eqref{L_5} is ill-defined~\cite{Schwartz1,Schwartz2} since the Dirac peak $\delta(y-L)$ enters in particular as a factor of the profiles $q^n_R(y)$ and $d^n_L(y)$ being not continuous
at $y=L$, as deduced from Eq.~\eqref{ELE_5integ} -- in the limit $\epsilon \to 0$ -- combined with the free BC imposing $d^n_R(L)\neq 0$, $q^n_L(L)\neq 0$ [see respectively 
the first and fourth solutions in Eq.~\eqref{completeEBCsm0}]~\footnote{The profiles $q^n_L(y)$, $d^n_R(y)$ are usually assumed to be continuous at $y=L-\epsilon$ while 
$q^n_R(y)$, $d^n_L(y)$ remain unknown exactly at this point.}.
Finally, the $q^n_R(y)$ and $d^n_L(y)$ jumps at $y=L$, obtained when regularising the brane-Higgs coupling, conflict with the field continuity axiom of the invoked  
theory of variation calculus and hence with the Hamilton's variational principle~\cite{HilbertBC}.

In the following steps of this Regularisation~I, one solves Eq.~\eqref{ELE_5ter} first in the interval $[0,L-\epsilon]$ (bulk EOM without Yukawa couplings) 
and applies the free BC at $y=0$ on the obtained profiles. Then one solves similarly Eq.~\eqref{ELE_5ter} on $[L-\epsilon,L]$ before applying the jump and continuity  
conditions~\eqref{ELE_5integ} at $y=L-\epsilon$ on the resulting profiles. The last step is to apply the free BC at $y=L$ on these profiles and take the limit $\epsilon \to 0$ 
(to recover the studied brane-Higgs model) on the written BC. The obtained BC give rise to the equation whose solutions constitute the fermion mass spectrum:
\begin{equation}
\tan(M_n \; L) = X \ \ (\forall n) \  .
\label{RegulResultI} 
\end{equation}
The absolute value of the mass spectrum induced by this equation is exactly the same as the 4D approach result of Eq.~\eqref{4D-prof-BC_Y_4}-\eqref{4D-prof-BC_Y_4bis}.

\subsubsection{Regularisation II drawbacks}

Within the Regularisation~II~\cite{Azatov:2009na,Casagrande:2010si,Barcelo:2014kha,Csaki:2003sh}, the Higgs coupling is first shifted in the bulk equations~\eqref{ELE_5} which become
\begin{equation}
\forall n\ , \ 
\left\{
\begin{array}{r c l}
\partial_4 q^n_R(y) + M_n \; q^n_L(y) &=& - \ \delta(y-[L-\epsilon]) \; X \; d^n_R(y) \ ,
\\ \vspace{-0.2cm} \\
\partial_4 q^n_L(y) - M_n \; q^n_R(y) &=& \delta(y-[L-\epsilon]) \; X' \; d^n_L(y) \ ,
\\ \vspace{-0.2cm} \\
\partial_4 d^n_R(y) + M_n \; d^n_L(y) &=& - \ \delta(y-[L-\epsilon]) \; X' \; q^n_R(y) \ ,
\\ \vspace{-0.2cm} \\
\partial_4 d^n_L(y) - M_n \; d^n_R(y) &=& \delta(y-[L-\epsilon]) \; X \; q^n_L(y) \ .
\end{array}
\right.
\label{ELE_5-shiftII}
\end{equation}
Integrating these four relations over an infinitesimal range centered at $y=L-\epsilon$ gives:
\begin{equation}
\forall n\ , \ 
\left\{
\begin{array}{r c l}
q^n_R([L-\epsilon]^+)-q^n_R([L-\epsilon]^-) &=& - \; X \; d^n_R(L-\epsilon) \ ,
\\ \vspace{-0.2cm} \\
q^n_L([L-\epsilon]^+)-q^n_L([L-\epsilon]^-)   &=& X' \; d^n_L(L-\epsilon) \ ,
\\ \vspace{-0.2cm} \\
d^n_R([L-\epsilon]^+)-d^n_R([L-\epsilon]^-)   &=& - \; X' \; q^n_R(L-\epsilon) \ ,
\\ \vspace{-0.2cm} \\
d^n_L([L-\epsilon]^+)-d^n_L([L-\epsilon]^-) &=& X \; q^n_L(L-\epsilon) \ .
\end{array}
\right.
\label{ELE_5integII}
\end{equation}
This set of conditions shows that the four wave functions undergo a jump at $y=L-\epsilon$ so that their derivative with respect to $y$ 
are not defined at this point. Hence the four integrations performed on Eq.~\eqref{ELE_5-shiftII} to obtain Eq.~\eqref{ELE_5integII} are not well defined in this regularisation. 
In other terms, the continuity conditions~\eqref{ELE_5integII} rely on the right-hand sides of the equations so that one must choose a value for each profile exactly at $y=L-\epsilon$.
Taking a standard mean value weighted thanks to a real number, $c$, Eq.~\eqref{ELE_5integII} takes the form
\begin{equation}
\forall n\ , \ 
\left\{
\begin{array}{r c l}
q^n_R([L-\epsilon]^+)-q^n_R([L-\epsilon]^-) &=& - \; X \; \frac{d^n_R([L-\epsilon]^-)\; + \; c \ d^n_R([L-\epsilon]^+)}{1+c} \ ,
\\ \vspace{-0.2cm} \\
q^n_L([L-\epsilon]^+)-q^n_L([L-\epsilon]^-)   &=& X' \; \frac{d^n_L([L-\epsilon]^-)\; + \; c\  d^n_L([L-\epsilon]^+)}{1+c}  \ ,
\\ \vspace{-0.2cm} \\
d^n_R([L-\epsilon]^+)-d^n_R([L-\epsilon]^-)   &=& - \; X' \; \frac{q^n_R([L-\epsilon]^-)\; + \; c\  q^n_R([L-\epsilon]^+)}{1+c}  \ ,
\\ \vspace{-0.2cm} \\
d^n_L([L-\epsilon]^+)-d^n_L([L-\epsilon]^-) &=& X \; \frac{q^n_L([L-\epsilon]^-)\; +\; c\  q^n_L([L-\epsilon]^+)}{1+c}  \ .
\end{array}
\right.
\label{ELE_5weight}
\end{equation}
Scrutinising the left-hand sides of those four equations, one observes that jumps may arise at $y=L$ (under the limit $\epsilon \to 0$) for the four profiles [for each 
excited $n^{th}$ mode]. Determining which profiles are discontinuous requires to consider the free BC at $y=L$ (before applying the limit $\epsilon \to 0$), 
the various $c$ values (including infinity) and the four profiles simultaneously [as they are related through Eq.~\eqref{ELE_5weight}].
The hypothesis that all of the four profiles are continuous at $y=L-\epsilon$ (in the limit $\epsilon \to 0$) corresponds to the same field configuration as in the 
absence of Yukawa interactions~\footnote{Free BC for continuous profiles and free version of the bulk equations~\eqref{ELE_5-shiftII} without the jump conditions~\eqref{ELE_5weight} 
at $y=L-\epsilon$ involving effectively the Yukawa couplings.} and leads thus to a free fermion mass spectrum. This kind of solution was not considered in the literature since it
does not reproduce the SM at low-energies and is thus not realistic. Therefore, there exists at least one profile discontinuous at $y=L$ which in turn cannot be derived at this point 
and leads to an undefined kinetic term [in the last two terms of 5D Action~\eqref{L_3}]. Furthermore, the obtained discontinuous [at $y=L$] profile comes in factor of $\delta(y-L)$ in 
Eq.~\eqref{L_5}, spoiling the mathematical validity of this action. Besides, once more, this jump obtained at $y=L$ within the regularisation process is not compatible with the field 
continuity axiom implicitly used when applying the Hamilton's variational principle.

In the next steps of Regularisation~II, Eq.~\eqref{ELE_5-shiftII} is first solved over the domain $[0,L-\epsilon]$ (free bulk EOM) 
and the free BC at $y=0$ are applied on the resulting wave functions. Eq.~\eqref{ELE_5-shiftII} is then solved over $[L-\epsilon,L]$ before the jump/continuity  
conditions~\eqref{ELE_5weight} at $y=L-\epsilon$ are applied on the obtained profiles. Finally the free BC at $y=L$ are implemented on those profiles and one applies 
the limit $\epsilon \to 0$ on the expressed BC. These BC make appear the following fermion mass spectrum equation for $c=1$:
\begin{equation}
\tan(M_n \; L) = \dfrac{4 X}{4 + X X'} \ \ [\forall n].
\label{RegulResultII}
\end{equation}

\subsection{Inconsistencies of the softened brane-Higgs coupling} 
\label{RegSoft}

Another type of regularisation used in the literature (on warped models)~\cite{Azatov:2009na,Casagrande:2010si,Carena:2012fk,Malm:2013jia,Hahn:2013nza,Barcelo:2014kha,Csaki:2003sh} 
consists in replacing the Dirac peak $\delta(y-L)$ of Eq.~\eqref{L_5} by a normalised square function which has a vanishing width ($\epsilon$) and an infinite value ($1/\epsilon$) in the limit $\epsilon \to 0$ 
where one expects to recover the considered model with a brane-Higgs coupling. Nevertheless, we point out here that the Dirac peak $\delta(y-L)$ at the Higgs brane, and in turn the 
original model, is not rigorously recovered via a limit, $\displaystyle{\delta(y-L)=\lim_{\epsilon \to 0}\eta_\epsilon(y-L)}$, of a so-called nascent delta function (or delta sequence) 
$\eta_\epsilon$ -- here the mentioned square function -- since such an equality is only symbolic: a distribution cannot be defined as the simple direct limit of a basic 
function~\footnote{Strictly speaking, it is the effect of the Dirac peak in the integration of a function $f(y)$ over an interval covering the point $y=L$, 
$\int \delta(y-L) f(y)dy=f(L)$, which can be reproduced via an integration of the type, ${\displaystyle{\lim_{\epsilon \to 0}}} \int \eta_\epsilon(y-L) f(y)dy=f(L)$, not 
performed in the present regularisation.}. Hence this would-be regularisation is not satisfactory in the sense that it does not strictly reproduce the studied brane-Higgs scenario. By the 
way, notice that no profile jump is needed to be imposed in this regularisation.

In addition, the two schemes of Regula\-risation~I and II still hold in this framework of a softened coupling and in case of Regula\-risation~I  
a problem arises again: some terms of the profile EOM are taken at zero based on the assumption that $0 \times \delta(0)=0$ whereas the quantity 
$0 \times \delta(0)$ is undefined.

\subsection{Two non-commutativities of calculation steps}

The analytical differences of the mass spectra found in the Regularisations~I and II, as well as via the softened and shifted brane-Higgs peaks, could be compensated by the different input values of the Yukawa coupling constant
parameters ($Y_5$ and $Y'_5$) to get identical physical mass values.
Nevertheless, the Regularisations~I and II are in fact physically different as induces the existence of measurable flavour violating effective 4D Yukawa couplings at leading order in $v^2/m_1^2$
which are generated by the $Y'_5$ couplings~\cite{Azatov:2009na} being present exclusively within Regularisation~II (as appears clearly in the 4D approach).
This physical difference between the two schemes of regularisation raises the paradoxal question, of which one is the sole correct analytical scheme to use, and represents thus as a confirmation 
of the inconsistency of regularising the Higgs peak.
These two schemes of regularisation are obtained~\cite{Barcelo:2014kha} by commuting in the 4D calculation (of masses and couplings) the ordering of implementation of the two limits $\epsilon \to 0$ 
[the regularising parameter $\epsilon$ defined in Eq.~\eqref{ELE_5ter}] and $N \to \infty$ [the upper value $N$ of the KK level $n$ in Eq.~\eqref{KK_1}]. 
Therefore, this physical non-commutativity of calculation steps reflects the inconsistency of the Higgs peak regularisation.
Another paradoxal non-commutativity of calculation steps arising in the context of regularisation of a brane-Higgs coupled to bulk fermions was discussed in Ref.~\cite{Carena:2012fk,Malm:2013jia}: 
different results of Higgs production/decay rates when taking $\epsilon \to 0$ and then $N_{KK} \to \infty$~\footnote{Here $N_{KK}$ 
stands for the number of excited fermion eigenstates exchanged at the loop-level.}~\cite{Casagrande:2010si} or the inverse order~\cite{Azatov:2010pf} in their calculation.
We can thus interpret now this second non-commutativity of calculation steps as being another effect, and in turn another confirmation, of the problematic Higgs regularisation (also expected with a warped extra dimension). 
Hence, the theoretical debate in the literature about the origins of those two non-commutativities [involving $\epsilon$] finds its solution in the mathematically ill-defined (see above) and unnecessary 
(see below) Higgs regularisation [introducing $\epsilon$].

\section{5D Treatment: the correct approach}
\label{Yukawa_terms_as boundary_conditions}

In this part, we consider the presence of the Yukawa couplings~\eqref{L_5} and present the rigorous 5D method to calculate the fermionic mass spectrum -- which does not require any kind of regularisation.  
We follow the main lines of the methodology developed for the free case in Section~\ref{free_bulk_fermions}.

\subsection{The naive approach}
\label{sub:naive}

For the fermion masses, the relevant part of the considered Action~\eqref{eq:actionTot} to start with is
\begin{eqnarray}
S^{\rm{m}}_{\rm{5D}} = S_{\Psi} + S_X - \int d^4x~dy~\delta(y)  \ \mathcal{L}_{B}  \ , \label{eq:actionNoH}
\end{eqnarray}
where the first term based on Eq.~\eqref{L_3} can be recast into Action~\eqref{modified_action_1} and $\mathcal{L}_{B}$ includes the BBT of Eq.~\eqref{eq:actionBound}. 
Regarding the free brane at $y=0$, we could equivalently apply the EBC~\eqref{courant_2} instead of including these BBT, as we have exposed in details in Sections~\eqref{sec:freeEBC}-\eqref{sec:freeNBC}.
Now without loss of generality, the least action principle leads to the four following conditions,
\begin{align}
0 =  \delta_{Q^\dagger_L} S^{\rm{m}}_{\rm{5D}} &= \displaystyle{ \int d^4x \;   dy \;  
\delta Q^\dagger_L \left[ i\bar{\sigma}^\mu \partial_\mu Q_L - \partial_4 Q_R  \right] } 
\nonumber \\ 
&\displaystyle{ + \int d^4x \left. \left[ \delta Q^\dagger_L \left( \dfrac{1}{2} Q_R -X D_R \right) \right] \right|_{L} - \int d^4x \ \left. \left( \delta Q^\dagger_L  Q_R \right)  \right|_{0} } \ ,
\nonumber  
\end{align}
\begin{align}
0 = \delta_{Q^\dagger_R} S^{\rm{m}}_{\rm{5D}} &= \displaystyle{ \int d^4x \; dy \;  
\delta Q^\dagger_R \left[ i\sigma^\mu \partial_\mu Q_R + \partial_4 Q_L  \right] } 
&\displaystyle{ + \int d^4x \left. \left[ - \delta Q^\dagger_R \left( \dfrac{1}{2} Q_L + X' D_L \right) \right] \right|_{L} } \ ,
\nonumber  
\end{align}
\begin{align}
0 =  \delta_{D^\dagger_L} S^{\rm{m}}_{\rm{5D}} &= \displaystyle{ \int d^4x \;   dy \;  
\delta D^\dagger_L \left[ i\bar{\sigma}^\mu \partial_\mu D_L - \partial_4 D_R  \right] } 
&\displaystyle{ + \int d^4x  \left. \left[ \delta D^\dagger_L \left( \dfrac{1}{2} D_R -X^{\prime \star} Q_R \right) \right] \right|_{L}  } \ ,
\nonumber  
\end{align}
\begin{align}
0 = \delta_{D^\dagger_R} S^{\rm{m}}_{\rm{5D}} & = \displaystyle{ \int d^4x \; dy \;  
\delta D^\dagger_R \left[ i\sigma^\mu \partial_\mu D_R + \partial_4 D_L  \right] } 
\nonumber \\ 
&\displaystyle{ + \int d^4x \left. \left[ -\delta D^\dagger_R \left( \dfrac{1}{2} D_L + X^{\star} Q_L \right) \right] \right|_{L} + \int d^4x \ \left. \left( \delta D^\dagger_R  D_L \right)  \right|_{0} } \ .
\label{eq:HVP_4relations}
\end{align}
Like in the studied free case, as the non-vanishing field variations $\delta F^\dagger_{L/R}$, $\left . \delta F^\dagger_{L/R} \right |_{0, L}$ are generic, 
the sum of the first two terms (first line), in each of those four relations, must vanish separately, which brings in the same equations as the 5D EOM~\eqref{ELE_1} 
and hence -- via the mixed KK decomposition~\eqref{KK_2} and 4D Dirac-Weyl equations~\eqref{Dirac_2} -- the profile equations
\begin{equation}
\forall n \geq 0 \, , \ 
\left\{
\begin{array}{r c l}
\partial_4 q_R^n(y) - M_n \, q_L^n(y) &=& 0
\\ \vspace{-0.2cm} \\
\partial_4 q_L^n(y) + M_n \, q_R^n(y) &=& 0
\\ \vspace{-0.2cm} \\
\partial_4 d_R^n(y) - M_n \, d_L^n(y) &=& 0
\\ \vspace{-0.2cm} \\
\partial_4 d_L^n(y) + M_n \, d_R^n(y) &=& 0
\end{array}
\right. \ 
\label{ELE_Yuk}
\end{equation}
whose solutions are found to be [with distinct constants for $f_{L/R}^n(y)=q_{L/R}^n(y)$ or $f_{L/R}^n(y)=d_{L/R}^n(y)$], as in Eq.~\eqref{profiles_rein}, 
\begin{equation}
f_{L}^n(y) = B^n_{R} \; \cos(M_n \; y) - B^n_{L} \; \sin(M_n \; y), \ \ 
f_{R}^n(y) = B^n_{L} \; \cos(M_n \; y) + B^n_{R} \; \sin(M_n \; y).
\label{profiles_reinM}
\end{equation}
The NBC resulting from Eq.~\eqref{eq:HVP_4relations} read as:
\begin{equation}
\left\{
\begin{array}{l}
\left. \left( Q_R - 2X \; D_R \right) \right|_{L} = 0,  \ \ \ \left. \left( D_R - 2X^{\prime \star}  \; Q_R \right) \right|_{L} = 0,   \ \ \ \left. Q_R \right|_0 = 0,  \\
\left. \left( Q_L + 2X' \; D_L \right) \right|_{L} = 0,  \ \ \ \left. \left( D_L + 2X^{\star}  \; Q_L \right) \right|_{L} = 0,  \ \ \ \left. D_L \right|_0 = 0. \\
\end{array}
\right. \ 
\label{BC_3}
\end{equation}
Combining these NBC leads to the following consistency conditions on the Lagrangian parameters,
\begin{equation}
4 X X^{\prime \star} = 4 X^\star X' = 1 \, ,
\label{XX'}
\end{equation}
and in turn to $4|XX'|=1$ and $\alpha_{Y'} = \alpha_Y +2 k \pi$ where $k$ is an integer and $X=\, \vert X\vert \, e^{i\alpha_Y}$, $X'=\, \vert X'\vert \, e^{i\alpha_{Y'}}$, 
the real numbers $\alpha_Y$, $\alpha_{Y'}$ representing the complex phases. The BC~\eqref{BC_3}, combined with the bulk profile EOM~\eqref{ELE_Yuk} [with 
solutions~\eqref{profiles_reinM}] taken at $y=L$, constitute the complete BC. Referring to the dependence on the quantity $X^{(\prime)}$, we denote $(\times)$ this new class of complete 
BC at the brane with a Yukawa coupling (here at $y=L$) to distinguish them from the Dirichlet BC usually noted $(-)$ or the Neumann BC noted $(+)$. The BC~\eqref{BC_3} on the 5D fields give rise to the following 
conditions on the profiles, through the KK decomposition~\eqref{KK_2}, 
\begin{equation}
\forall n \geq 0 \, , \ \left\{
\begin{array}{l}
  q_R^n(L) - 2X \; d_R^n(L)   = 0,  \ \ \   d_R^n(L) - 2X^{\prime \star}  \; q_R^n(L)     = 0,   \ \ \  q_R^n(0)  = 0,  \\
  q_L^n(L) + 2X' \; d_L^n(L)    = 0,  \ \ \   d_L^n(L) + 2X^{\star}  \; q_L^n(L)    = 0,  \ \ \ d_L^n(0)  = 0, \\
\end{array}
\right. \ 
\label{BC_4}
\end{equation}
since the 4D fermion fields for the mass eigenstates cannot be linearly related -- as discussed below Eq.~\eqref{ELE_2}.
Those profile conditions, once applied on the solutions~\eqref{profiles_reinM}, lead to the form,
\begin{equation}
\forall n \geq 0 \, , \ \left\{
\begin{array}{l}
q_{L}^n(y) = C^n_R \, \cos(M_n \, y),  \ \ \ q_{R}^n(y) =  C^n_R \, \sin(M_n \, y), \\
d_{L}^n(y) =  - D^n_L \, \sin(M_n \, y),  \ \ \ d_{R}^n(y) = D^n_L \, \cos(M_n \, y),
\end{array}
\right. \ 
\label{prof-BC_Y_1}
\end{equation}
together with the relations,
\begin{eqnarray}
\tan (M_n \, L) = 2X \, \dfrac{D_L^n}{C_R^n} = 2X^\star \, \dfrac{C_R^n}{D_L^n} \ \ \Rightarrow \ \ \tan^2(M_n \, L) = 4|X|^2 \, , 
\nonumber\\
\cot (M_n \, L) = 2X' \, \dfrac{D_L^n}{C_R^n} = 2X^{\prime \star} \, \dfrac{C_R^n}{D_L^n} \ \ \Rightarrow \ \ \cot^2(M_n \, L) = 4|X^{\prime}|^2 \, .
\label{syst_BC_1}
\end{eqnarray}
These last two mass spectrum relations induced are strictly equivalent thanks to Eq.~\eqref{XX'}. The obtained mass spectrum allows to determine
for instance the BC $(-\times)$ of the profile $d_{L}^n(y)$: $d_{L}^n(0)=0$ and $d_{L}^n(L)=- D^n_L \, \sin(M_n \, L)$.

Let us check the validity of the obtained solutions. In the decoupling limit of high KK masses (compared to the typical SM energy scale) applied to the present model, 
one expects to recover approximately the SM set-up at low-energies. 
This decoupling condition is necessary for the theoretical consistency of the model and it is generically imposed by the experimental constraints. 
First, according to Eq.~\eqref{syst_BC_1}, the lightest mode mass is,
\begin{align}
M_0 &= \dfrac{1}{L} \arctan(\pm 2|X|) =  \dfrac{1}{L} \arctan(\pm \sqrt{2}|Y_4Lv|) \underset{m_1 \gg |v| }{\sim}  \pm\sqrt{2} |Y_4v| \, , 
\label{decoupling_mass}
\end{align}
since $m_1=\pi/L$ [{\it c.f.} Eq.~\eqref{PossSolSpctr}]. This 4D effective fermion mass [{\it c.f.} Eq.~\eqref{Dirac_2}] is well proportional to the Higgs VEV as in the SM.
Secondly, the effective Yukawa coupling constant in the 4D action term involving the lightest modes, $-\int d^4x \, Y_{00}H\psi^{0\dagger}_L\psi^0_R + \rm{H.c.}$, 
is obtained by injecting the KK decompositions \eqref{KK_2} in Eq.~\eqref{eq:actionYuk} and then integrating over $y$, by using the wave functions~\eqref{prof-BC_Y_1} 
to take into account the mass mixings induced by the Yukawa couplings (5D method):
\begin{align}
Y_{00} &= \dfrac{Y_5}{L} \, q_L^{0\star}(L) \, d_R^0(L) + \dfrac{Y_5^{\prime\star}}{L} \, d_L^{0\star}(L) \, q_R^0(L)  \nonumber \\ 
 &= \dfrac{Y_5}{L} \, C_R^{0\star} D_L^0 \, \cos^2(M_0 \, L) -  \dfrac{Y_5^{\prime\star}}{L} \, D_L^{0\star} C_R^0 \, \sin^2(M_0 \, L)  \nonumber \\
 &=  \dfrac{Y_5}{L} \, C_R^{0\star} D_L^0 \, \cos^2(M_0 \, L) -  4 \dfrac{Y_5^{\prime\star}}{L} \, D_L^{0\star} C_R^0 \, XX^\star \cos^2(M_0 \, L) \nonumber \\
  &=  \dfrac{Y_5}{L} \, C_R^{0\star} D_L^0 \, \cos^2(M_0 \, L) -  4 \dfrac{Y_5^{\prime\star}}{L} \, \frac{(C_R^{0\star})^2}{D_L^{0\star}}C_R^0 \, X^2 \cos^2(M_0 \, L) \nonumber \\
  &=  \dfrac{Y_5}{L} \, C_R^{0\star}  \, \cos^2(M_0 \, L) \left [ D_L^0 -  \frac{C_R^{0\star}}{D_L^{0\star}}C_R^0 \right ]
 \underset{m_1 \gg |v| }{=} 0 \, ,  
\label{decoupling_Yuk}
\end{align}
where we have used subsequently the deduced equation and the relation involving $X^\star$ in the first line of Eq.~\eqref{syst_BC_1} before invoking Eq.~\eqref{XX'};
as indicated right above, for high KK mass values, $C_R^0C_R^{0\star}=D_L^0D_L^{0\star}$ so that $Y_{00}$ vanishes which differs from the SM framework.  
Indeed, applying the ortho-normalisation condition~\eqref{normalization_2}, for $n=m=0$, to the solution profiles~\eqref{prof-BC_Y_1}, we deduce that 
\begin{eqnarray}
\int_0^{L} dy \; \left[ q^{0\star}_L(y) \; q^0_L(y) + d^{0\star}_L(y) \; d^0_L(y) \right] & = & \int_0^{L} dy \; \left[ q^{0\star}_R(y) \; q^0_R(y) + d^{0\star}_R(y) \; d^0_R(y) \right] 
\nonumber \\ \Leftrightarrow \ 
\vert C_R^0 \, \vert^2 \,  \frac{\sin(2M_0 L)}{2M_0 L} & = & \vert D_L^0 \, \vert^2 \,  \frac{\sin(2M_0 L)}{2M_0 L} 
\label{eq:cond_CD}
\end{eqnarray}
which induces $\vert C_R^0 \vert^2 = \vert D_L^0 \vert^2$ (and in turn $\vert D_L^0 \vert^2\neq 0$) 
in the decoupling limit of Eq.~\eqref{decoupling_mass} where $\vert 2M_0 L \vert \underset{m_1 \gg |v| }{\sim} 2\sqrt{2} |Y_4vL| \ll 2\sqrt{2} |Y_4| \pi = {\cal O}(\pi)$ 
and hence $\vert 2M_0 L \vert<\pi$ so that we can divide Eq.~\eqref{eq:cond_CD} by $\sin(2M_0 L)/2M_0 L$ being non-vanishing~\footnote{We can also justify that $\sin(2M_0 L)/2M_0 L$ is not vanishing from 
the deduced relation in the first line of Eq.~\eqref{syst_BC_1} since one needs $\vert X\vert \neq 0$ to have $M_0\neq 0$ [{\it c.f.} Eq.~\eqref{decoupling_mass}] when $ m_1 \gg |v| $ (decoupling condition on the 
fermion mass).}. 
The decoupling condition is thus not respected which reveals a problem in the present treatment of the studied model. The problematic 
vanishing of the effective 4D Yukawa coupling constant $Y_{00}$ results from the invariance of the 
Action~\eqref{eq:actionNoH} under the exchange transformation, $Q \leftrightarrow D$ together with $Y^\star_5 \leftrightarrow Y'_5$ at $y=L$ [symmetry also explicit in Eq.~\eqref{ELE_Yuk} and \eqref{BC_4}]: 
this symmetry will be broken in the correct treatments presented below.
A confirmation of the failure of the present 5D treatment is the non-matching of the obtained spectrum equation~\eqref{syst_BC_1} with the 4D matrix method result~\eqref{4D_spectrum}.
Therefore, the treatment of the brane-Higgs coupling of this subsection should be reconsidered: we present the other methods in the next two subsections.

\subsection{Introducing the fermion current condition [EBC]}
\label{sub:CurrentCond}

Like in the free case treated in Section~\ref{sec:freeEBC}, we now try to define well the geometrical field configuration of the considered scenario based on the action 
$S^{\rm{m}}_{\rm{5D}}$ of Eq.~\eqref{eq:actionNoH}. 
In this scenario, the two 5D fields $Q,D$ propagate only inside the interval ${\cal C}^1 \equiv [0,L]$. This set-up translates into a condition 
of vanishing probability current at both boundaries. The current is here the sum of the two individual currents of type~\eqref{courant_1} for the two species $Q,D$ since those 
fermions are mixed together through the mass terms~\eqref{L_5}. To find out this current form rigorously, we first vary the action as in the beginning of Section~\ref{sub:naive} 
and deduce the 5D EOM~\eqref{ELE_1} whose profile solutions were given in Eq.~\eqref{profiles_reinM}.
Then using the obtained EOM~\eqref{ELE_1}, we apply in Appendix~\ref{Noether_THM} the Noether theorem
to work out the probability current~\eqref{currentGENYuk}~\footnote{This result holds as well in the case without BBT.} which reads as, 
\begin{equation}
j^M = - \alpha \left ( \bar{Q} \; \Gamma^M Q + \bar{D} \; \Gamma^M D \right ) \, ,
\label{courant_1Y}
\end{equation}
as dictated by the global $U(1)$ symmetry of the Action~\eqref{eq:actionNoH} relying on the transformations,
\begin{equation}
Q_{L/R} \mapsto \text{e}^{i \alpha} Q_{L/R} \ , \ D_{L/R} \mapsto \text{e}^{i \alpha} D_{L/R}  \ .
\label{symetrie_1Y} 
\end{equation}
$\alpha$ ($\in \mathbb{R}$) is a continuous parameter [now forced by the invariant terms~\eqref{L_5} to be common for the two fields $Q,D$]
involved for example in the infinitesimal field varia\-tions,  
\begin{equation}
\underline{\delta} Q_L = i \alpha Q_L \, , \, \underline{\delta} Q^\dagger_L = - i \alpha Q^\dagger_L \ .
\label{var-Yuk} 
\end{equation}
We thus find that the effect of the Yukawa interactions 
is not to modify the currents but rather to force one to add them up for having a probability conservation relation (due to the induced mixing among the $Q$ and $D$ fields).
Finally, the condition of vanishing probability current at the boundary where is located the Yukawa coupling reads as~\footnote{The current condition at the other boundary is taken into account 
through the BBT in the last term of Eq.~\eqref{eq:actionNoH}.},
\begin{equation}
\left. j^4 \right |_{ L}= - \alpha \left. \left ( \bar{Q} \; \Gamma^4 Q + \bar{D} \; \Gamma^4 D \right ) \right |_{L} 
= i \alpha \left. \left( Q^\dagger_R Q_L - Q^\dagger_L Q_R + D^\dagger_R D_L - D^\dagger_L D_R \right) \right|_{L} = 0 \ .
\label{courant_2Y}
\end{equation}
For a non-trivial transformation with $\alpha \neq 0$, the field variation of this relation is 
\begin{equation}
\left. \left(
\delta Q_R^\dagger Q_L + Q_R^\dagger \delta Q_L
- \delta Q_L^\dagger Q_R - Q_L^\dagger \delta Q_R
+ \delta D_R^\dagger D_L + D_R^\dagger \delta D_L
- \delta D_L^\dagger D_R - D_L^\dagger \delta D_R
\right) \right|_{L} = 0 .
\label{current_variations_0}
\end{equation}
The variation calculus chronology here is quite simple as no field is fixed by the EBC~\eqref{courant_2Y}: the fields [and their respective variations] are instead related via this
Eq.~\eqref{courant_2Y} [and Eq.~\eqref{current_variations_0}]. 
Now the part of the variation of the action $S^{\rm{m}}_{\rm{5D}}$, from Eq.~\eqref{eq:actionNoH}, containing the boundary terms is written in Eq.~\eqref{eq:App-noEBC} of the Appendix~\ref{BC-EBC}. 
The complementary variation of the bulk action vanishing separately was already used just above to derive the 5D EOM~\eqref{ELE_1}. Notice that this variation of the bulk action with respect to 
the non-conjugate 5D fields in $\delta_{F_{L,R}} S^{\rm{m}}_{\rm{5D}}$ requires an integration by part to recover the Hermitian conjugate form of the EOM~\eqref{ELE_1} [visible in Eq.~\eqref{eq:HVP_4relations}] 
and the boundary terms in $\delta_{F_{L,R}} S^{\rm{m}}_{\rm{5D}}$ [visible in Eq.~\eqref{eq:App-noEBC}].  
One could think of obtaining NBC and their Hermitian conjugate form respectively from $\delta_{F_{L,R}} S^{\rm{m}}_{\rm{5D}}$ and $\delta_{F^\dagger_{L,R}} S^{\rm{m}}_{\rm{5D}}$ [as obtained in Eq.~\eqref{BC_3}], 
in Eq.~\eqref{eq:App-noEBC}, but in fact all the field variations are connected via the relation~\eqref{current_variations_0} so that one can not get rid of those directly. 
There is no consistent way of combining the NBC~\eqref{eq:App-noEBC} [even by splitting it into several vanishing expressions] with the EBC~\eqref{current_variations_0} in order to get some set of 
BC and another set made of the Hermitian conjugate BC, except in the special but excluded case (see right below) where,
\begin{equation}
Q_L|_{L}=D_L|_{L}=Q_R|_{L}=D_R|_{L}=0 \, .
\label{DrastBC}
\end{equation}
One could impose the condition~\eqref{eq:App-noEBC} to be realised separately leading 
to the NBC~\eqref{BC_3} (and their Hermitian conjugate form) which induce~\footnote{As can be seen by replacing $Q^\dagger_{L,R}|_{L}$ and $D_{L,R}|_{L}$ in the expression~\eqref{courant_2Y} thanks to the two relations
in the first and second column of Eq.~\eqref{BC_3} respectively.} the EBC~\eqref{courant_2Y}, but one would then be back to the case of Section~\ref{sub:naive}
which has been ruled out due to the decoupling limit argue and the non-matching of the 4D versus 5D results.
The drastic BC~\eqref{DrastBC} (or the Hermitian conjugate form), imply obviously both the EBC~\eqref{courant_2Y} and NBC~\eqref{eq:App-noEBC} 
but lead to an inconsistency which reveals a problematic solution. Indeed, the BC~\eqref{DrastBC}, once applied to the solutions~\eqref{profiles_reinM}, induce two equations that, after being squared and summed together,
give the identity $(B^n_{R})^2=(B^n_{L})^2$ [$\forall n\geq 0$] for both the $Q$ and $D$ fields: considering any $n$-level,
the case $B^n_{R}=\pm B^n_{L}=0$ (for $Q$ and $D$) conflicts with the normalisation condition~\eqref{normalization_2}, while for 
$B^n_{R}=\pm B^n_{L}\neq 0$ (for at least one of the two fields $Q,D$) the two mentioned equations result in the simultaneous equalities $\cos(M_nL)=\sin(M_nL)$ and $\cos(M_nL)= - \sin(M_nL)$ whose unique solution
$\cos(M_nL)=\sin(M_nL)=0$ makes no sense. 
As a conclusion, the impossibility, to combine the EBC and NBC for getting some set of BC together with a complementary set made of their Hermitian conjugate, conflicts with the Feynman prescription for particles and anti-particles
-- according to which the fields and their Hermitian conjugate undergo identical physical equations (up to complex conjugate coupling constants). 
This conflict~\footnote{As described in the free case [below Eq.~\eqref{modified_action_1}], the direct injection of the EBC~\eqref{courant_2Y} in the Action~\eqref{eq:actionNoH} would cancel out the boundary 
terms at $y=L$ in Eq.~\eqref{modified_action_1} spoiling then the needed hermiticity of the whole action $S^{\rm{m}}_{\rm{5D}}$ and leading thus to a related problem.} 
shows that the present approach of the configuration with a Yukawa coupling located at a boundary, based on the vanishing of the fermion current taken as an EBC, is not yet the correct one.
The origin of the problem being that the current~\eqref{courant_1Y} does not contain an explicit term that involves the Yukawa coupling constant.

\subsection{Introducing the bilinear brane terms [NBC]}
\label{ActVar}

As in the free case, we try here to apply the alternative treatment, based on the introduction of the BBT at $y=L$, in order to develop a consistent approach.
We consider the fermion part of the Action~\eqref{eq:actionTot}:
\begin{eqnarray}
S^{\rm{\prime m}}_{\rm{5D}} = S_{\Psi}  + S_B + S_X \ , \label{eq:actionNoHcomp}
\end{eqnarray}
based on the kinetic part~\eqref{modified_action_1}, the BBT~\eqref{eq:actionBound} and the mass terms~\eqref{L_5}.
The boundary fields $\left . F \right |_{0, L}$ are initially unknown so that their functional varia\-tions will be taken non-vanishing:
$\left . \delta F \right |_{0, L}  \neq 0$. Without loss of generality, the stationary action condition can be split into the two following conditions for each field $F=Q,D$
[extending Eq.~\eqref{HVP_3a}-\eqref{HVP_3b} to include the Yukawa terms] together with the two other equations $\delta_{F_L} S^{\rm{\prime m}}_{\rm{5D}}=\delta_{F_R} S^{\rm{\prime m}}_{\rm{5D}}=0$, 
\begin{eqnarray}
0 =  \delta_{F^\dagger_L} S^{\rm{\prime m}}_{\rm{5D}} & = & \displaystyle{ \int d^4x \; \int_0^{L} dy \;  
\left[ \delta F^\dagger_L \; i\bar{\sigma}^\mu \partial_\mu F_L - \delta F^\dagger_L \; \partial_4 F_R  \right] } 
\ \label{HVP_4aa} \\
& & \displaystyle{ + \int d^4x \left \{ \; C^L_F \, \left[ \left . \left( \delta F^\dagger_L \; F_R \right) \right |_{L} - \left . \left( \delta F^\dagger_L \; F_R \right) \right |_{0} \; \right]  \right . }
\nonumber \\ 
& & \displaystyle{ \left . - C^L_F \, X \left . \left( \delta {F}^\dagger_L \; D_R \right) \right |_{L} - C^R_F \, X^{\prime \star} \left . \left( \delta {F}^\dagger_L \; Q_R \right) \right |_{L} \right \} } \ ,
\nonumber \\
0 = \delta_{F^\dagger_R} S^{\rm{\prime m}}_{\rm{5D}} & = & \displaystyle{ \int d^4x \; \int_0^{L} dy \;  
\left[ \delta F^\dagger_R \; i\sigma^\mu \partial_\mu F_R + \delta F^\dagger_R \; \partial_4 F_L  \right] } 
\ \label{HVP_4bb} \\
& & \displaystyle{ + \int d^4x \left \{ \; C^R_F \,  \left[ - \left . \left( \delta F^\dagger_R \; F_L \right) \right |_{L} + \left . \left( \delta F^\dagger_R \; F_L \right) \right |_{0} \; \right]  \right . }
\nonumber \\ 
& & \displaystyle{ \left . - C^R_F \, X^{\star} \left . \left( \delta {F}^\dagger_R \; Q_L \right) \right |_{L} - C^L_F \, X^{\prime} \left . \left( \delta {F}^\dagger_R \; D_L \right) \right |_{L} \right \} } \ ,
\nonumber 
\end{eqnarray}
using the same $C^{L,R}_{Q,D}$ definitions as in Eq.~\eqref{HVP_3a}-\eqref{HVP_3b}. 
Once more, the non-vanishing field variations $\delta F^\dagger_{L/R}$, $\left . \delta F^\dagger_{L/R} \right |_{0, L}$ being generic, 
the sum of the first two terms (first line) in Eq.~\eqref{HVP_4aa} and \eqref{HVP_4bb} respectively must vanish separately, which brings in the 5D EOM~\eqref{ELE_1} 
and in turn -- through the mixed KK decomposition~\eqref{KK_2} and 4D Dirac-Weyl equations~\eqref{Dirac_2} -- the wave function equations~\eqref{ELE_Yuk}
with solutions as in Eq.~\eqref{profiles_reinM}:
\begin{eqnarray}
q_{L}^n(y) = B^n_{R} \; \cos(M_n \; y) - B^n_{L} \; \sin(M_n \; y), \ \ q_{R}^n(y) = B^n_{L} \; \cos(M_n \; y) + B^n_{R} \; \sin(M_n \; y), \ \ \ \ \
\label{profiles_Y} \\
d_{L}^n(y) = D^n_{R} \; \cos(M_n \; y) - D^n_{L} \; \sin(M_n \; y), \ \ d_{R}^n(y) = D^n_{L} \; \cos(M_n \; y) + D^n_{R} \; \sin(M_n \; y), \ \ \ \ \
\nonumber
\end{eqnarray}
using here new constant parameters $B^n_{L/R}$, $D^n_{L/R}$. Note that, in contrast, 
grouping directly the terms involving $C^{L/R}_F$ factors with the first two terms in Eq.~\eqref{HVP_4aa} and \eqref{HVP_4bb} -- simply thanks to the introduction of Dirac peaks at $y=0,L$ -- 
would lead to mathematically meaningless relations between 5D functions and distributions. Finally, the NBC resulting from Eq.~\eqref{HVP_4aa}-\eqref{HVP_4bb} read as:
\begin{eqnarray}
& \left. \left( Q_R - X \; D_R \right) \right|_{L} = 0, \ \ \ \left. \left( D_L + X^{\star}  \; Q_L \right) \right|_{L} = 0, \ \ \ X^{\prime \star} \left . Q_R \right|_L = 0, \ \ \ X'  \left. D_L \right|_L = 0 ,
\nonumber \\ 
&   \left. Q_R  \right |_{0}    = 0 , \ \ \ \left .  D_L \right |_{0}  = 0.
\label{BC_13G}
\end{eqnarray}
Integrating by part the bulk terms in the other relations $\delta_{F_L} S^{\rm{\prime m}}_{\rm{5D}}=\delta_{F_R} S^{\rm{\prime m}}_{\rm{5D}}=0$ 
allows to recover the Hermitian conjugate form of the EOM~\eqref{ELE_1} as well as the Hermitian conjugate form of the NBC~\eqref{BC_13G}. 
The NBC~\eqref{BC_13G} can be rewritten without loss of generality as,
\begin{eqnarray}
& \left. \left( Q_R - X \; D_R \right) \right|_{L} = 0, \ \ \ \left. \left( D_L + X^{\star}  \; Q_L \right) \right|_{L} = 0, \ \ \ 
X^{\prime}=0 \ {\rm or} \ \{ \left . Q_R \right|_L = 0 \ {\rm and}  \left. D_L \right|_L = 0 \},
\nonumber \\ 
&   \left. Q_R  \right |_{0}    = 0 , \ \ \ \left .  D_L \right |_{0}  = 0,
\nonumber 
\end{eqnarray}
and in turn as,
\begin{eqnarray}
{\rm \underline{BC \ 1:}} & \left. X \; D_R \right|_{L} = 0, \ \ \ \left. X^{\star}  \; Q_L \right|_{L} = 0, \ \ \ 
\left . Q_R \right|_L = 0, \ \ \ \ \left. D_L \right|_L = 0 , \ \ \ 
   \left. Q_R  \right |_{0}    = 0 , \ \ \ \left .  D_L \right |_{0}  = 0,
\nonumber \\ 
{\rm or\, ,}
\nonumber \\ 
{\rm \underline{BC \ 2:}} & \left. \left( Q_R - X \; D_R \right) \right|_{L} = 0, \ \ \ \left. \left( D_L + X^{\star}  \; Q_L \right) \right|_{L} = 0, \ \ \ 
X^{\prime}=0,  \ \ \  \left. Q_R  \right |_{0}    = 0 , \ \ \ \left .  D_L \right |_{0}  = 0.
\nonumber \\ \label{BC_13ter}
\end{eqnarray}
The lightest fermionic state possesses a mass equal to the $\alpha_{00}$ element of the 4D mass matrix~\eqref{M} in the decoupling limit $m_1 \to \infty$ of the studied high-energy scenario, 
which allows to reproduce well the SM mass expression at the low-energy scales. For this purpose, one must have in particular a non-vanishing Yukawa coupling constant and $X \neq 0$ so that the BC~1 read as, 
\begin{eqnarray}
{\rm \underline{BC \ 1:}} & \left. D_R \right|_{L} = 0, \ \ \ \left. Q_L \right|_{L} = 0, \ \ \ 
\left . Q_R \right|_L = 0, \ \ \ \ \left. D_L \right|_L = 0 ,  \ \ \ \
 \left. Q_R  \right |_{0}    = 0 , \ \ \ \left .  D_L \right |_{0}  = 0,
\nonumber 
\end{eqnarray} 
BC at $y=L$ exactly similar to those in Eq.~\eqref{DrastBC} which have been ruled out. Hence we exclude the BC~1. 
Let us move to the BC~2 which can be expressed in terms of the profiles, thanks to the relevant mixed KK 
decomposition~\eqref{KK_2}, as follows (together with the condition $X'=0$),
\begin{eqnarray}
{\rm \underline{BC \ 2:}} \ &\forall n\geq 0 \, , \ q^n_R(L) - X \, d^n_R(L)  = 0, \ d^n_L(L) + X^{\star} \, q^n_L(L)  = 0,  \   q^n_R(0)    = 0 , \  d^n_L(0) = 0 \ .
\nonumber 
\end{eqnarray}
So these BC~2 at $y=0$ applied on the solutions~\eqref{profiles_Y} produce the following profiles, 
\begin{eqnarray}
q_{L}^n(y) = B^n_{R} \; \cos(M_n \; y)  \ ,
\ \ q_{R}^n(y) =  B^n_{R} \; \sin(M_n \; y) \ ,
\label{prof-BC_Y_1bis}
\\ d_{L}^n(y) =  - D^n_{L} \; \sin(M_n \; y) \ ,
\ \ d_{R}^n(y) = D^n_{L} \; \cos(M_n \; y)  \ . 
\nonumber
\end{eqnarray}
One must be careful to avoid some of the mathematical inconsistencies also encountered in the regularisation procedures of Section~\ref{Usual_treatment}: 
in particular the existence of any profile jump at the interval boundaries which would induce an undefined derivative term in the 5D Action based on Eq.~\eqref{L_3} 
[last two terms], an ill-defined term in the Action~\eqref{L_5} -- where the Dirac peak $\delta(y-L)$ would come in factor of a profile discontinuous at $y=L$  --
and finally would conflict with the field continuity axiom of the invoked theory of variation calculus. Therefore, we are taking all the profiles continuous at both 
boundaries which is the reason why we have applied the BC~2 at $y=0$ on the bulk expressions~\eqref{profiles_Y}. 
The application of the BC~2 at $y=L$ on the resulting bulk expressions~\eqref{prof-BC_Y_1bis} gives rise to the relations
[using $M_n, \, B^n_{R} , \, D^n_{L} \, \neq 0, \forall n\geq 0$, to be checked a posteriori],
\begin{eqnarray}
\tan(M_n \; L) = X \frac{D^n_{L}}{B^n_{R}} = X^{\star} \frac{B^n_{R}}{D^n_{L}} \  ,
\nonumber
\end{eqnarray}
which can be recast into (via $X D^n_{L} \hat = \vert X \vert D^{\prime n}_{L}$, $\forall n\geq 0$)
\begin{eqnarray}
\tan(M_n \; L) = \pm \, \vert X \vert, \ D^{\prime n}_{L}= \pm \, B^n_{R} \  ,
\label{prof-BC_Y_3}
\end{eqnarray}
assuming that the generic phase $\alpha_Y$ of the 5D Yukawa coupling constant, defined by  
$X=\vert X \vert \, e^{i\alpha_Y}$ as below Eq.~\eqref{XX'}, is included into a new parameter $D^{\prime n}_{L}\hat =D^n_{L}\, e^{i\alpha_Y}
= \vert D^n_{L}\vert \, e^{i(\alpha_0+\alpha_Y)}$. At this level, it is important to highlight the fact that it is the same $\pm$ sign entering the two equalities in 
Eq.~\eqref{prof-BC_Y_3}. We already remark the real mass spectrum resulting from Eq.~\eqref{prof-BC_Y_3}, even for a Yukawa coupling constant with 
a non-vanishing imaginary part. Now let us first apply the orthogonality conditions of Eq.~\eqref{normalization_2} on the solutions~\eqref{prof-BC_Y_1bis}: 
\begin{eqnarray}
\int_0^{L} dy \, \left[ D^{n\star}_{L} D^m_{L} \, \sin(M_n \, y) \sin(M_m \, y) + \, B^{n\star}_{R} B^m_{R} \cos(M_n \, y) \cos(M_m \, y) \right] = 0, \ \forall n\neq m \nonumber \\ 
\int_0^{L} dy \, \left[ B^{n\star}_{R} B^m_{R} \, \sin(M_n \, y) \sin(M_m \, y) + D^{n\star}_{L} D^m_{L} \, \cos(M_n \, y) \cos(M_m \, y) \right] = 0.  \ \ \ \  \label{orthogonal_Y} 
\end{eqnarray}
We insert trigonometric formulae~\footnote{Of the kind, $\cos(M_n \, y) \cos(M_m \, y) = [ \cos(M_n \, y+M_m \, y) + \cos(M_n \, y-M_m \, y) ] /2 $.} in this equalities, in order to 
perform the integration, and then make use of another type of trigonometric relation~\footnote{$\ \sin(M_n \, L - M_m \, L) = \sin(M_n \, L) \cos(M_m \, L) - \cos(M_n \, L) \sin(M_m \, L)$.} 
to obtain the following simplified form [$\forall n, m \ {\rm with} \ n \neq m$],
\begin{eqnarray}
\frac{\sin(M_m L) \cos(M_n L)}{M_n^2-M_m^2} \left [ M_m \, B^{n\star}_{R} B^m_{R} + M_n \, D^{n\star}_{L} D^m_{L} \right ] 
=& \left [ M_n \, B^{n\star}_{R} B^m_{R} + M_m \, D^{n\star}_{L} D^m_{L} \right ]
\nonumber \\
& \times \ \frac{\sin(M_n L) \cos(M_m L)}{M_n^2-M_m^2}  ,
\nonumber \\
\frac{\sin(M_m L) \cos(M_n L)}{M_n^2-M_m^2} \left [ M_m \, D^{n\star}_{L} D^m_{L} + M_n \, B^{n\star}_{R} B^m_{R} \right ] 
= & \left [ M_n \, D^{n\star}_{L} D^m_{L} + M_m \, B^{n\star}_{R} B^m_{R} \right ]
\nonumber \\
&   \times \ \frac{\sin(M_n L) \cos(M_m L)}{M_n^2-M_m^2}  . 
\nonumber
\end{eqnarray}
We can divide these equalities by $\cos(M_n L)\cos(M_m L)$ since $\cos(M_n L)\neq 0$ [$\forall n$] (as the mass spectrum given by Eq.~\eqref{prof-BC_Y_3}
is not a free one); we get [$\forall n, m \ {\rm with} \ n \neq m$],
\begin{eqnarray}
\tan(M_m L) \left [ M_m \, B^{n\star}_{R} B^m_{R} + M_n \, D^{n\star}_{L} D^m_{L} \right ] 
= 
\tan(M_n L) \left [ M_n \, B^{n\star}_{R} B^m_{R} + M_m \, D^{n\star}_{L} D^m_{L} \right ], 
\ \nonumber \\
\tan(M_m L) \left [ M_m \, D^{n\star}_{L} D^m_{L} + M_n \, B^{n\star}_{R} B^m_{R} \right ] 
= 
\tan(M_n L) \left [ M_n \, D^{n\star}_{L} D^m_{L} + M_m \, B^{n\star}_{R} B^m_{R} \right ].
\ \nonumber 
\end{eqnarray}
Using the spectrum of Eq.~\eqref{prof-BC_Y_3}, getting rid of the common factor $(M_n-M_m)$ in those two equations  
and dividing the resulting equalities by the constant parame\-ters allowing to separate the $n$ and $m$ dependences, we find the unique relation
\begin{eqnarray}
\frac{D^{n\star}_{L}}{B^{n\star}_{R}}  = \frac{B^{m}_{R}}{D^{m}_{L}},  \  \forall n, m \ {\rm with} \ n \neq m,
\nonumber
\end{eqnarray}
which is clearly true since we know from Eq.~\eqref{prof-BC_Y_3} that $B^{n}_{R}=\pm D^{\prime n}_{L}=\pm D^{n}_{L}\, e^{i\alpha_Y}$ ($\forall n$). 
Secondly, we apply the normalisation conditions of Eq.~\eqref{normalization_2} on the profile solutions~\eqref{prof-BC_Y_1bis} obeying the constraints~\eqref{prof-BC_Y_3}
[implying $\vert B^{n}_{R} \vert ^2=\vert D^{n}_{L} \vert ^2$]:
\begin{eqnarray}
\int_0^{L} dy \, \vert B^{n}_{R} \vert ^2 \, \left[ \sin^2(M_n \, y) +  \cos^2(M_n \, y)   \right] = L, \  \forall n. \nonumber
\label{normalization_Y} 
\end{eqnarray}
Those conditions allow to complete Eq.~\eqref{prof-BC_Y_3} which becomes,
\begin{eqnarray}
\forall n \, , \ 
\tan(M_n \; L) = \pm \, \vert X \vert, \   D^{\prime n}_{L}= \pm \, B^n_{R},  \ \vert B^{n}_{R} \vert =\vert D^{n}_{L} \vert = 1 \ ,
\label{prof-BC_Y_3new}    
\end{eqnarray}
and exhibits then the three following types of solutions,
\begin{eqnarray} 
& {\rm I:} \ \forall n, \ \tan(M_n \; L) = + \, \vert X \vert, \ B^{n}_{R} = e^{i(\alpha_0+\alpha_Y)}, 
\ D^{n}_{L} = e^{i\alpha_0}, \nonumber
\\ 
& {\rm II:} \ \forall n, \ \tan(M_n \; L) = - \, \vert X \vert, \ B^{n}_{R} = e^{i(\alpha_0+\alpha_Y\pm\pi)}, 
\ D^{n}_{L} = e^{i\alpha_0}, \nonumber
\\
& {\rm III: \ Solution \ I \ for \ some \ {\it n} \ values \ and \ II \ for \ other \ {\it n} \ values},  \nonumber
\end{eqnarray}
$\alpha_0$ becoming the common phase (defined modulo $2\pi$). Following a similar discussion as the one below Eq.~\eqref{4D-prof-BC_Y_4}-\eqref{4D-prof-BC_Y_4bis},
we can claim that for the whole set of $n$-levels the absolute value of the fermion mass has the following common expression for the Solutions~I and II: 
\begin{eqnarray}
\tan(M_n \; L) = + \vert X \vert & \Rightarrow & \vert M_n \vert = \fbox{$ \left \vert \dfrac{\arctan(\vert X \vert) + (-1)^n \, \tilde n(n) \, \pi}{L} \right \vert $} \ , \label{prof-BC_Y_4} \\
{\rm \ or, \ \ } \tan(M_n \; L) = - \vert X \vert & \Rightarrow & \vert M_n \vert \equiv \fbox{$ \left \vert \dfrac{\arctan(\vert X \vert) + (-1)^n \, \tilde n(n) \, \pi}{L} \right \vert $} \ ,
\nonumber
\end{eqnarray}
using the $\tilde n(n)$ function already defined in Eq.~\eqref{ntilde}. Once again, 
the Solution~III must be excluded as the complete and consistent infinite mass spectrum solution is determined within a unique model hypothesis selected among the two 
given mathematical solutions, I or II, since Eq.~\eqref{prof-BC_Y_4} fixes the quantity $M_n L$ modulo $\pi$.

Within the simplified case of a real 5D Yukawa coupling constant 
($\vert X \vert =X$), we thus find that the unique tower~\eqref{prof-BC_Y_4} of absolute values of the physical fermion masses is matching 
the one obtained in the 4D approach: Eq.~\eqref{4D-prof-BC_Y_4}-\eqref{4D-prof-BC_Y_4bis}. 
This exact 4D-5D matching confirms the overall consistency of our calculations -- without regularisations -- 
and is of course expected to be reached as well for a complex 5D Yukawa coupling constant.

In particular, the insensitivity of the 4D fermion mass matrix~\eqref{M} to the $Y'_5$ coupling constant [described below Eq.~\eqref{coeff}] 
matches interestingly the spectrum independence on $Y'_5$ induced by the result $Y'_5=0$ obtained in the BC~2 [see Eq.~\eqref{BC_13ter}] used for the 5D point of view.

Let us give an intuitive interpretation of the absence of r\^ole for the $Y_5'$ coupling (involved in $X'$) in the final spectrum~\eqref{prof-BC_Y_4} which
depends only on the $X$ quantity. Starting with the free Action $S_{\Psi} + S_B$, the profiles $d^n_L(y)$ and $q^n_R(y)$ [$\forall n$], defined by Eq.~\eqref{KK_1} 
and with solutions~\eqref{completeEBCsm0}, vanish in particular at the boundary $y=L$. 
Hence the term with a $X'$ coefficient in the Action piece $S_X$ of Eq.~\eqref{L_5}, once added to $S_{\Psi} + S_B$, is expected to have a vanishing 
factor coming from the integration over the interval due to the Dirac peak. This argument is only intuitive as it does not really include
the possible `back reaction' effect of the $X'$ term on the profiles via modified BC.

Finally, let us discuss the condition on the fermion current for the boundary at $y=L$.
Inserting the four expressions of $Q^{(\dagger)}_R$ and $Q^{(\dagger)}_L$, provided by the first two NBC of Eq.~\eqref{BC_13ter} at $y=L$, in the current condition~\eqref{courant_2Y} leads to a trivially 
true equality (all the four terms involving exclusively $D^{(\dagger)}_{L,R}$ fields cancel each other). 
This feature means that the NBC~\eqref{BC_13ter} imply the condition~\eqref{courant_2Y} so that the geometrical field set-up of the present model with matter stuck on an interval is well defined.

As a conclusion, adding BBT at the brane with the Yukawa coupling to bulk fermions permits a consistent treatment of the considered scenario and a correct calculation of the mass spectrum.

\section{Overview and implications}
\label{sec:over}

\subsection{The action content}

In Table~1, we summarise the results for the obtained fermion BC at a single 3-brane.
We conclude from this table that for fermions on an interval and coupled or not to a brane-localised Higgs field, either BBT should be generated in the action 
or conditions should arise on the fermion current (forcing then the 4D treatment in case of a brane Yukawa coupling) depending on the origin of the model at high-energies. 
In the same spirit, notice that the UV completion will determine whether the selection of fermion boundary conditions is imposed or deduced from the action form. 
The UV completion should not bring simultaneously EBC (imposing vanishing currents) and BBT (guaranteeing current vanishing) because it would be possible but redundant.
It is interesting to observe anyway that the necessary additional fermionic ingredient, with respect to the kinetic terms, reveals that limiting the integration domain of the action does not suffice 
to define consistently and completely the basic field configuration along the interval (or more generally over a compactified space).
Indeed, without having a vanishing fermion current at a boundary, one could imagine a source of creation or a mechanism of absorption for particles at the boundary.
Therefore, the present status, resulting from this analysis and its synthesis, is that the action expression may not contain all the information ({\it e.g.} current conditions) 
needed to define an higher-dimensional model, regarding the geometrical set-up and field configurations.

$$ 
 \begin{array}{c|c|c|c}
  &  \ \  \mbox{\bf No boundary} \ \  & \  \mbox{\bf Vanishing current }  &  \ \mbox{\bf Bilinear brane}      \\
    &  \ \  \mbox{\bf characteristic}  \ \  & \   \mbox{\bf condition [EBC]}   & \  \mbox{\bf terms [NBC]}     \\
& & &  \\
\hline
& & &  \\
\mbox{{\bf 4D approach}}  \ \    &  \mbox{\it (Impossible)} & \mbox{BC ($\pm$)}  & \mbox{BC ($\pm$)} \\
& & &  \\
 \hline
& & &  \\
\mbox{{\bf 5D approach}}  \ \   & \mbox{\it (Impossible)} &  \mbox{\it (Impossible)} &  \mbox{BC ($\times$)} \\
 & & &  \\
  \end{array} 
$$ 
\begin{center}
 {\small \underline{Table~1:} Bulk fermion BC (when a consistent determination exists) at a 3-brane where is located the Higgs boson coupled to these fermions, for different boundary treatments:
presence of BBT, vanishing of the probability current or nothing specific.
The 4D line holds as well for the 5D approach of the free brane. As usual, the Dirichlet BC are noted $(-)$, the Neumann BC $(+)$ and we denote  
$(\times)$ the new BC depending on the Yukawa coupling constant. See also the main text for the BC, NBC and EBC acronym definitions.}
\end{center}

\vspace{0.5cm}

\begin{figure}[!h]
\begin{center}
\vspace{0.5cm}
\includegraphics[width=10cm]{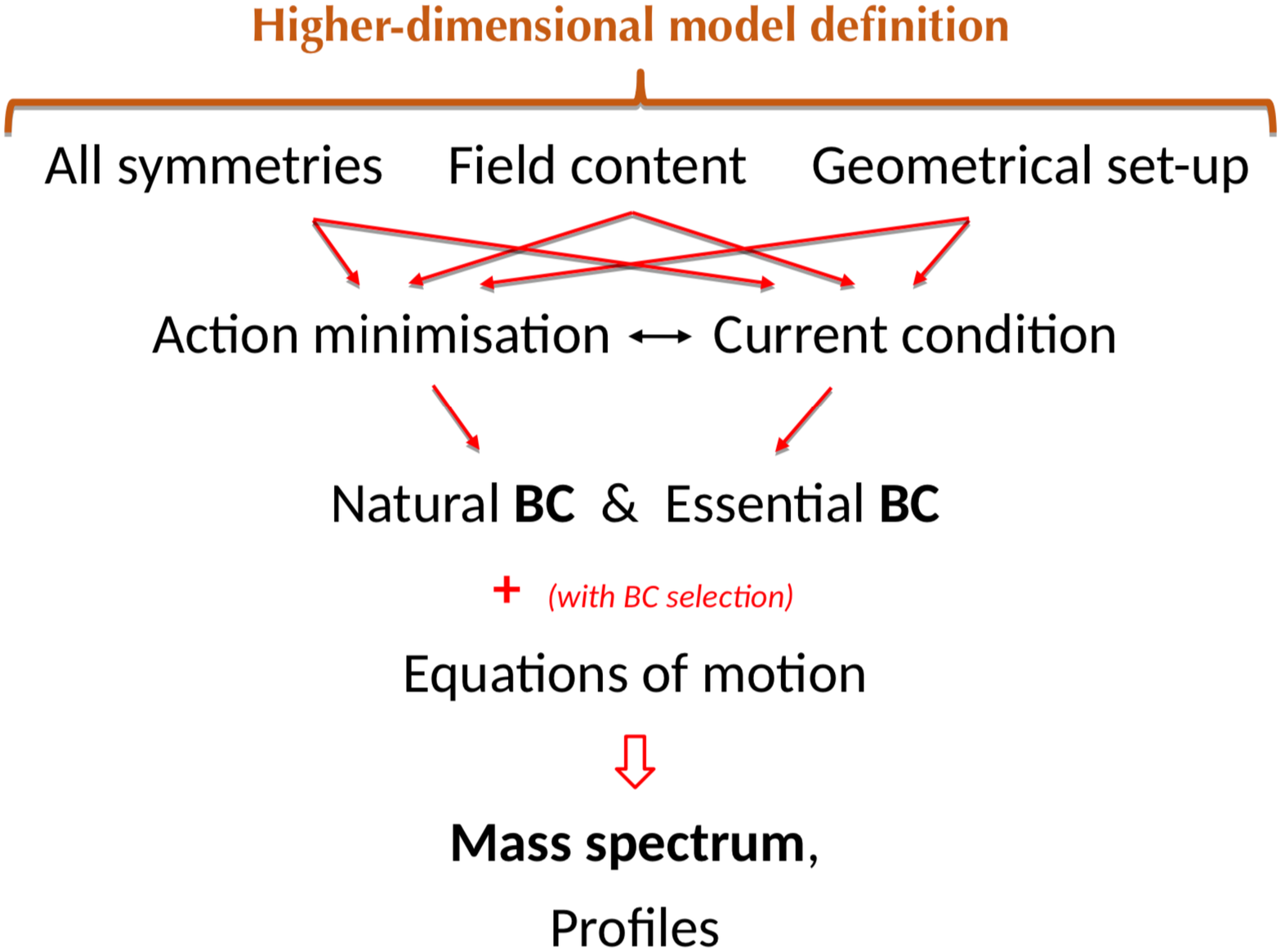}
\vspace{-0.5cm}
\end{center}
\caption{
Inverse pyramidal picture illustrating the general principles for determining the wave functions and masses of mixed KK modes within a given model based on extra dimension(s).
Same notations as in the main text are used. 
}\label{fig:Pyramidal}
\end{figure}

Based on the above results, we describe now the generic methodology to find out the mass spectrum and KK wave functions 
(allowing to calculate the 4D effective couplings) along the extra spatial dimension(s) of a given scenario. For this purpose, we present in 
Fig.~\ref{fig:Pyramidal} a schematic description of the main principles. The figure must be understood as follows. 
A given extra-dimensional model must be first defined by its geometrical set-up [space-time structure and field location configuration], 
its field content and its internal (gauge groups,\dots) as well as other types (the Poincar\'e group here) of symmetries. 
These three types of informations determine entirely the action form~\footnote{Within a well-understood scenario, all terms of the Lagrangian density should be deduced from the model definition 
exclusively: absence of couplings from symmetries, presence of  
BBT from the geometrical set-up, etc} whose minimisation gives rise to the 5D EOM and NBC. Besides the geometrical hypotheses of the model, concerning for instance the space limits for field propagation, 
may produce probability current conditions translating into EBC~\footnote{The EBC could also originate from the definition of the symmetry of orbifold scenarios.} 
which must be combined with these NBC. At this level, a choice of the combined BC obtained can be required
(if not determined automatically by the action structure itself). Then the KK decomposition (together with the EOM on the 4D fields) allows to derive the EOM and BC on the KK
profiles along the extra dimension(s). The last step is obviously to solve these profile EOM, coupled to the complete BC, in order to work out the mass spectra.

\subsection{Implementation of the cut-off on energy}

We have to discuss the cut-off treatment as the framework of higher-dimensional models is non-renormalisable theories which are valid in a limited domain of energy, up to a certain scale, set by perturbativity 
conditions on effective dimensionless couplings. If the UV completion of such models affects the KK excitation towers and thus the fermion mass spectrum, in an unknown way, then its calculation must includes the KK
state masses only up to the cut-off value typically (the UV corrections at low-energies can be parametrised via higher-dimensional operators). In a case of absence of UV effects on the specific mass spectrum sector, the whole 
KK towers should be taken into account at the mass calculation level since even the smallest mass eigenvalues can be affected by the mixing effects of the infinite towers. Now in both situations, only the eigenstates with masses 
up to the cut-off scale should be considered for the phenomenological observables (reaction amplitudes, rates,...) 
due to the non-renormalisable nature of the theory. Technically, the implementation of an energy cut-off in the bulk fermion mass calculation and tree-level Lagrangian construction forces the use of the 4D approach.
Indeed, the mixed KK decomposition~\eqref{KK_2}, used in the 5D approach, includes the mixing of the whole tower: the fields $\psi_{L,R}^n ( x_\mu)$ are mass eigenstates.

\subsection{Phenomenological impacts}

In the appropriate treatment developed in the present paper, without regularisation, the obtained mass spectrum and effective 4D Yukawa coupling depend on $Y_5$ but not on the $Y'_5$ coupling constant. 
For instance in Eq.~\eqref{decoupling_Yuk}, one should in fact apply the result $Y'_5= 0$ as dictated by the relevant BC~2 in Eq.~\eqref{BC_13ter}.
Applying an energy cut-off in the process of mass calculation would not affect this independence on $Y'_5$ as is clear from the point of view of the 4D approach.

The results for fermion masses and profiles are also correct when one invokes the Higgs peak Regularisation~I which cancels out the $Y'_5$ dependence. 
Hence, the phenomenological analyses of the literature based on such results are still valid: see for instance 
Ref.~\cite{Huber:2000ie,Huber:2003sf,Casagrande:2008hr,Goertz:2008vr,Moreau:2005kz,Moreau:2006np,Ledroit:2007ik,Bouchart:2009vq,Gopalakrishna:2013hua}. 
Those studies apply on the geometrical background with warped extra dimensions where
the KK spectrum independence on $Y'_5$ is expected to occur as well.

Note that the results from Regularisation~I and the correct ones, in the approach without regularisation at all, are exactly identical only by accident. Indeed in the Regularisation~I, 
considering first the 5D treatment, the mass spectrum calculation in presence of Yukawa couplings suffers from two errors which exactly cancel out each other:
there are no BBT, which affects the resulting spectrum equation by a factor $2$ [as seen when comparing the spectra with BBT in Eq.~\eqref{prof-BC_Y_4} and without BBT in Eq.~\eqref{syst_BC_1}], 
and a regularisation is applied. Now starting from the 4D treatment of Regularisation~I and adding BBT (or current conditions) would have no effect on the 4D mass matrix [as described in Section~\ref{brTermSec}] 
like avoiding the regularisation process [as there is no analytical effect of Regularisation~I in which the limit $\epsilon\to 0$ is taken at the first step~\cite{Barcelo:2014kha}]: the results of Regularisation~I
are thus the same as in the correct approach.

In contrast, if the Higgs peak Regularisation~II is used, the obtained fermion masses and 4D Yukawa couplings depend on both $Y_5$ and $Y'_5$ so that the results differ effectively from the correct ones. 
Hence, the phenomenological studies based on these analytical results (for example Ref.~\cite{Azatov:2009na,Casagrande:2010si,Carena:2012fk,Malm:2013jia,Hahn:2013nza}) should be reconsidered or redone. 

For example, the effective 4D Yukawa couplings to fermions and their KK excitations affect the main Higgs production mechanism at the Large Hadron Collider (LHC): the gluon-gluon fusion via triangular loops of (KK) fermions.
Hence the effect of the realistic limit~\cite{Barcelo:2014kha} of vanishing $Y'_5$ on the constraints on KK masses derived in the studies~\cite{Casagrande:2010si,Carena:2012fk,Malm:2013jia,Hahn:2013nza}, 
within the warped background and based on the Regularisation~II, should be calculated precisely.

Besides, the rotation matrices diagonalising the 4D fermion mass matrix~\eqref{M} do not diagonalise simultaneously the effective 4D Yukawa coupling matrix 
since the last one does not contain matrix elements made of the pure KK masses. 
The induced flavour violating 4D Yukawa couplings are generated at leading order by $Y'_5$ contributions as can be shown diagrammatically~\cite{Azatov:2009na}.
Hence there exist large Flavour Changing Neutral Current (FCNC) effects in measured $\Delta F= 2$ processes such as $\bar K-K$, $\bar B-B$ and $\bar D-D$ mixings, mainly produced by tree-level exchanges of the Higgs boson 
via $Y'_5$ couplings, which lead to considerable lower bounds on the KK boson mass scale (in balance via opposite Yukawa coupling dependences with the ones from the tree-level contribution of the 
KK gluon exchange) found to be around $6-9$~TeV in the analysis~\cite{Azatov:2009na} 
on warped extra dimensions using indeed the Regularisation~II. Hence these bounds should be significantly suppressed in the realistic situation where $Y'_5\to 0$;
this limit should indeed be applied since the independence found in the present paper upon $Y'_5$ (extended via flavour indices) remains true for the case of three flavours, as well as for fermion bulk masses, as is clear in the 
4D approach where the $\beta_{ij}$-elements~\eqref{coeff} of the mass matrix still vanish. The predictions of Ref~\cite{Azatov:2009na}, based on Regularisation~II, that FCNC reactions involving Yukawa couplings, like the rare top 
quark decay $t\to ch$ and exotic Higgs boson decay to charged leptons $h\to \mu \tau$, can be observable at the LHC deserve reconsiderations as well when $Y'_5=0$.

\section{Summary and conclusions}

For bulk fermions coupled to a brane-Higgs boson, we have shown that the proper calculation of the fermion masses and effective 4D Yukawa couplings does not rely on Higgs peak regularisations.
The justifications are the following ones: {\it (i)} There are no fermion wave function jumps at the Higgs boundary so no motivation to introduce an arbitrary regularisation, {\it (ii)} the regularisations
suffer from several mathematical discrepancies confirmed by two known non-commutativities of calculation steps, {\it (iii)} the right method without any regularisation is validated in particular by the 
converging results of the 4D versus 5D treatments.

In the rigorous method developed for both free and brane-coupled bulk fermions, 
we have also pointed out the necessity to either include BBT in the Lagrangian density, or alternatively impose vanishing conditions for probability currents at the interval boundaries. 
Here the arguments go as follows: {\it (i)} the presence of BBT guarantees the vanishing current conditions which define the field geometrical configuration of the model, {\it (ii)} the BBT and
current conditions allow to find physically consistent fermion masses, bulk profiles and effective 4D Yukawa couplings (solutions fulfilling the normalisation constraints, the Hermitian conjugate BC 
and the decoupling limit condition), {\it (iii)} the BBT lead to the expected matching between the 4D and 5D calculation results.

The BBT represent a possible origin of the chiral nature of the SM as well as of its chirality distribution among quark/lepton $SU(2)_L$ doublets and singlets. 
Those terms could thus provide new clues about the UV completion of the SM. 

Depending on the UV completion, the general methodology worked out reveals that the informations regarding the definition of an higher-dimensional model are not necessarily fully contained in the action itself
-- through the deduced EOM and NBC -- but might be partly included as well in the EBC.

We have finished the analysis by the descriptions of the appropriate energy cut-off procedure in the present framework, and, of the phenomenological impacts of the new calculation method which predicts
the independence of the fermion masses and effective 4D Yukawa couplings on the $Y_5'$ parameter of the Lagrangian. This different coupling feature, with respect to the Regularisation~II usually 
applied in the literature, should in particular suppress significantly the previously obtained severe bounds on KK masses induced by FCNC processes generated via flavour violating couplings of the 
Higgs boson.

An extension of the present study, to generic BBT, fermion bulk masses, warped extra dimensions and orbifold scenarios, is under progress~\cite{WIP}.
\\ \\ \\ \noindent
{\bf Acknowledgments} \\ 
{\it The authors thank Emilian Dudas, Ulrich Ellwanger, Adam Falkowski, Maria Kitsara and J\'er\'emie Quevillon for useful discussions.
R.L. is supported by the agreement signed between the China Scholarship Council (CSC) and the University Paris-Sud.
F.N. acknowledges support from the IDEX Paris-Saclay. G.M. and F.N. are grateful to the 
"Commission du coll\`ege doctoral" of the University Paris-Saclay as well as to the University Paris-Sud for the complementary funding.}

\newpage

\appendix

\section*{Appendix}

\section{Notations \& conventions}
\label{notations_and_conventions}

Throughout the present paper, we use the conventions of the Ref.~\cite{Schwartz:2013pla}.
\vspace{0.3cm}\\
The 5D Minkowski metric is,
\begin{equation}
\eta_{MN} = \text{diag}(+1, -1, -1, -1, -1).
\label{metric_1}
\end{equation}
The 4D Dirac matrices are taken in the Weyl representation,
\begin{equation}
\gamma^\mu =
\begin{pmatrix}
0 & \sigma^\mu \\
\bar{\sigma}^\mu & 0
\end{pmatrix}
\phantom{000} \text{with} \phantom{000}
\left\{
\begin{array}{r c l}
\sigma^\mu &=& \left( \mathbb{I}, \sigma^i \right), \\
\bar{\sigma}^\mu &=& \left( \mathbb{I}, -\sigma^i \right),
\end{array}
\right.
\label{gamma_1}
\end{equation}
where $\sigma^i$ ($i = 1, 2, 3$) are the three Pauli matrices:
\begin{equation}
\sigma^1 =
\begin{pmatrix}
0 & 1 \\
1 & 0
\end{pmatrix},
\phantom{000}
\sigma^2 =
\begin{pmatrix}
0 & -i \\
i & 0
\end{pmatrix},
\phantom{000}
\sigma^3 =
\begin{pmatrix}
1 & 0 \\
0 & -1
\end{pmatrix} .
\end{equation}
One has also the 4D chirality operator,
\begin{equation}
\gamma^5 = i \prod_{\mu = 0}^3 \gamma^\mu =
\begin{pmatrix}
- \mathbb{I} & 0 \\
0 & \mathbb{I}
\end{pmatrix}.
\label{gamma_2}
\end{equation}
With our conventions, the 5D Dirac matrices read as,
\begin{equation}
\Gamma^M = \left( \gamma^\mu, i \gamma^5 \right).
\label{gamma_3}
\end{equation}

\section{Noether theorem including brane-localised terms}
\label{Noether_THM}

Here we demonstrate the Noether theorem in the presence of boundary-localised Yukawa couplings and BBT.
We first consider the free Action~\eqref{eq:actionKin} together with the BBT~\eqref{eq:actionBound} [or \eqref{eq:actionBoundCUSTO}] being invariant under the transformations~\eqref{symetrie_1} affecting 
the fields but not the coordinates $z^M$. The infinitesimal action variation under such a transformation on the field $F$ reads generically as, 
\begin{eqnarray}
\underline{\delta} (S_\Psi+S_{\rm B}) =  
\int d^4x~\left \{ -  \ \left . \underline{\delta} F^\alpha \frac{\partial  \mathcal{L}_{\rm B}}{\partial F^\alpha} \right \vert_0 \ 
-  \ \left . \underline{\delta} \bar F^\alpha \frac{\partial  \mathcal{L}_{\rm B}}{\partial \bar F^\alpha} \right \vert_0 \
+ \ \left . \underline{\delta} F^\alpha \frac{\partial  \mathcal{L}_{\rm B}}{\partial F^\alpha} \right \vert_L  
+ \ \left . \underline{\delta} \bar F^\alpha \frac{\partial  \mathcal{L}_{\rm B}}{\partial \bar F^\alpha} \right \vert_L \right \} 
\nonumber \\  + 
\int d^4x~dy~\left \{ \underline{\delta} F^\alpha \frac{\partial \mathcal{L}_{\Psi}}{\partial F^\alpha} + \underline{\delta} \bar F^\alpha \frac{\partial \mathcal{L}_{\Psi}}{\partial \bar F^\alpha} 
+ \, \underline{\delta} (\partial_M F^\alpha) \frac{\partial \mathcal{L}_{\Psi}}{\partial ( \partial_M F^\alpha)} +
\underline{\delta} (\partial_M \bar F^\alpha) \frac{\partial \mathcal{L}_{\Psi}}{\partial ( \partial_M \bar F^\alpha)} \right \} 
 \ . \nonumber \\
\label{deltabarS-1}
\end{eqnarray}
Now we invoke the generic version of the EOM, $\frac{\partial \mathcal{L}_{\Psi}}{\partial F^\alpha} =  \partial_M \frac{\partial \mathcal{L}_{\Psi}}{\partial (\partial_MF^\alpha)}$, as found in 
Eq.~\eqref{ELE_1}~\footnote{Of course similar EOM hold for the complex conjugate fields.},  
not including the possible BBT contributions rather involved in the separate NBC, as found in Eq.~\eqref{BC_1}-\eqref{complete_boundary_term} [without BBT] and \eqref{HVP_4n}-\eqref{HVP_4ncusto} [with BBT].
Using these EOM to rewrite the first two terms in the second line of Eq.~\eqref{deltabarS-1} and then grouping those with the last two terms to make global derivatives appear, we find:
\begin{eqnarray}
\underline{\delta} (S_\Psi+S_{\rm B}) =  
\int d^4x~\left \{ -  \ \left . \underline{\delta} F^\alpha \frac{\partial  \mathcal{L}_{\rm B}}{\partial F^\alpha} \right \vert_0 \ 
-  \ \left . \underline{\delta} \bar F^\alpha \frac{\partial  \mathcal{L}_{\rm B}}{\partial \bar F^\alpha} \right \vert_0 \
+ \ \left . \underline{\delta} F^\alpha \frac{\partial  \mathcal{L}_{\rm B}}{\partial F^\alpha} \right \vert_L  
+ \ \left . \underline{\delta} \bar F^\alpha \frac{\partial  \mathcal{L}_{\rm B}}{\partial \bar F^\alpha} \right \vert_L \right \} 
\nonumber \\  +  
\int d^4x~dy~\left \{ 
\, \partial_M \left ( \underline{\delta} F^\alpha \frac{\partial \mathcal{L}_{\Psi}}{\partial ( \partial_M F^\alpha)} \right ) +
\partial_M \left ( \underline{\delta} \bar F^\alpha \frac{\partial \mathcal{L}_{\Psi}}{\partial ( \partial_M \bar F^\alpha)} \right ) \right \} 
 \ . \nonumber \\
\label{deltabarS-2}
\end{eqnarray}
The four terms in the first line (right-hand side) of this equation vanish since the infinite\-simal field variations~\eqref{delta-bar-free} lead for instance to,
\begin{equation}
- \left . \underline{\delta} Q^\alpha \frac{\partial  \mathcal{L}_{\rm B}}{\partial Q^\alpha} \right \vert_0 
-  \ \left . \underline{\delta} \bar Q^\alpha \frac{\partial  \mathcal{L}_{\rm B}}{\partial \bar Q^\alpha} \right \vert_0 \
 \, =   \ \left . \frac{1}{2} \bar Q (i\alpha Q)  \right \vert_0 \, + \, \left . \frac{1}{2} (-i\alpha \bar Q) Q  \right \vert_0 \, =  \, 0 \ .
\label{d-brane-vanish} 
\end{equation}
A similar cancellation, due to the symmetry of the model, arises for the last two terms at $y=L$ and the $D$ field contributions (relying on the $\alpha'$ parameter).

The infinitesimal variation of the invariant Lagrangian from Eq.~\eqref{eq:actionKin} and \eqref{eq:actionBound} vanishes when integrated over the whole space [$\underline{\delta} (S_\psi+S_B)=0$] 
and even over any 5D domain $\Omega$, as the transformation affects the fields only.
The first line (right-hand side) of Eq.~\eqref{deltabarS-2} vanishes as well for any integration volume $\Omega$ due to relations of type~\eqref{d-brane-vanish} when $\Omega$ includes the boundaries
$y=0,L$ and due to the absence of Dirac peak in the integration domain in the other case.
Therefore, mathematically, Eq.~\eqref{deltabarS-2} implies the vanishing of its second line for any integration region $\Omega$ and in turn 
(the fields being fixed by the geometrical model configuration over the whole interval) the local conservation relation for the 5D probability current of the field $F$, 
\begin{equation}
\phantom{0} \partial_M j_F^M = 0 \, , \  \text{for any $z^M$} , \  \text{with} \ 
j_F^M \, = \, \underline{\delta} F^\alpha \frac{\partial \mathcal{L}_{\Psi}}{\partial ( \partial_M F^\alpha)} +
 \underline{\delta} \bar F^\alpha \frac{\partial \mathcal{L}_{\Psi}}{\partial ( \partial_M \bar F^\alpha)}  \  .
\label{currentGEN}
\end{equation}

Note that an alternative instructive reading, based on the global derivatives of the second line in Eq.~\eqref{deltabarS-2} and an integration over a generic 5D domain $\Omega$, is that the second line vanishing   
leads to a cancellation of the sum over the differences of current components (each difference integrated over the complementary dimensions). 
This cancellation expresses the 5-current conservation over all directions (equality of the global ingoing and outgoing currents with respect to a given hyper-volume $\Omega$) and is thus nothing else but a strictly 
equivalent, and less convenient, form of the conservation relation~\eqref{currentGEN}: global versus local conservation of the full current $j^M_F$. For a consistency check, let us wonder what happens when
the entire 5D domain is considered ({\it i.e.} $\Omega$ represents the whole 5D bulk). Then the differences $j_F^\mu(+\infty,y)-j_F^\mu(-\infty,y)$ tend to zero -- due to the vanishing of fields at infinite coordinates imposed 
by the vanishing boundary terms issued from the least action principle and {\it independently} to the wave function normalisation conditions -- so that one gets $\int d^4x \, [j_F^4(x^\mu,L)-j_F^4(x^\mu,0)]=0$.  
This specific conservation property of the 5-current (or of the matter presence probability) must be compatible with the geometrical field configuration defining the model. The definition of the interval,
$j_F^4(x^\mu,L)=j_F^4(x^\mu,0)=0$ [$\forall x^\mu$] as in Eq.~\eqref{courant_2} for the present scenario, satisfies well this conservation property. It is obviously not the only way to respect the property. 
For example, within an orbifold scenario, the boundary point identification establishing the space periodicity, $y=0 \equiv L \Rightarrow j_F^4(x^\mu,0) = j_F^4(x^\mu,L)$, realises as well the mentioned conservation pattern.

Let us now extend the demonstration of the Noether theorem to the presence of BBT and boundary-localised Yukawa couplings by
considering the free Action~\eqref{eq:actionKin} together with the BBT~\eqref{eq:actionBound} and the Yukawa terms~\eqref{eq:actionYuk}.
This whole action $S_\Psi+S_{\rm B}+S_{\rm Y}$ is invariant under the transformation~\eqref{symetrie_1Y}.
The infinitesimal action variation under this transformation reads as, 
\begin{eqnarray}
& & \underline{\delta} (S_\Psi+S_{\rm B}+S_{\rm Y})  =  \sum_{F=Q_{L/R},D_{L/R}} \, 
\int d^4x~\left \{ -  \ \left . \underline{\delta} F^\alpha \frac{\partial  \mathcal{L}_{\rm Y}}{\partial F^\alpha} \right \vert_L 
-  \ \left . \underline{\delta} F^{\dagger\alpha} \frac{\partial  \mathcal{L}_{\rm Y}}{\partial F^{\dagger\alpha}} \right \vert_L
\ \right \} 
\nonumber \\  & & + \sum_{F=Q,D} \, 
\int d^4x~\left \{ -  \ \left . \underline{\delta} F^\alpha \frac{\partial  \mathcal{L}_{\rm B}}{\partial F^\alpha} \right \vert_0 \ 
-  \ \left . \underline{\delta} \bar F^\alpha \frac{\partial  \mathcal{L}_{\rm B}}{\partial \bar F^\alpha} \right \vert_0 \
+ \ \left . \underline{\delta} F^\alpha \frac{\partial  \mathcal{L}_{\rm B}}{\partial F^\alpha} \right \vert_L  
+ \ \left . \underline{\delta} \bar F^\alpha \frac{\partial  \mathcal{L}_{\rm B}}{\partial \bar F^\alpha} \right \vert_L \right \} 
\nonumber \\  & & + \sum_{F=Q,D} \, 
\int d^4x~dy~\left \{ \underline{\delta} F^\alpha \frac{\partial \mathcal{L}_{\Psi}}{\partial F^\alpha} + \underline{\delta} \bar F^\alpha \frac{\partial \mathcal{L}_{\Psi}}{\partial \bar F^\alpha} 
+ \, \underline{\delta} (\partial_M F^\alpha) \frac{\partial \mathcal{L}_{\Psi}}{\partial ( \partial_M F^\alpha)} +
\underline{\delta} (\partial_M \bar F^\alpha) \frac{\partial \mathcal{L}_{\Psi}}{\partial ( \partial_M \bar F^\alpha)} \right \} 
 \ . \nonumber \\
\label{deltabarS-1Yuk}
\end{eqnarray}
Invoking once more the EOM, $\frac{\partial \mathcal{L}_{\Psi}}{\partial F^\alpha} =  \partial_M \frac{\partial \mathcal{L}_{\Psi}}{\partial (\partial_MF^\alpha)}$,  
including neither the possible BBT contributions nor the Yukawa terms (both rather entering the separate NBC),
we can rewrite the first two terms in the third line of Eq.~\eqref{deltabarS-1Yuk} and then grouping those with the last two terms to make global derivatives arise:
\begin{eqnarray}
& & \underline{\delta} (S_\Psi+S_{\rm B}+S_{\rm Y})  =  \sum_{F=Q_{L/R},D_{L/R}}  \, 
\int d^4x~\left \{ -  \ \left . \underline{\delta} F^\alpha \frac{\partial  \mathcal{L}_{\rm Y}}{\partial F^\alpha} \right \vert_L 
-  \ \left . \underline{\delta} F^{\dagger\alpha} \frac{\partial  \mathcal{L}_{\rm Y}}{\partial F^{\dagger\alpha}} \right \vert_L
\ \right \} 
\nonumber \\  & & + \sum_{F=Q,D} \, 
\int d^4x~\left \{ -  \ \left . \underline{\delta} F^\alpha \frac{\partial  \mathcal{L}_{\rm B}}{\partial F^\alpha} \right \vert_0 \ 
-  \ \left . \underline{\delta} \bar F^\alpha \frac{\partial  \mathcal{L}_{\rm B}}{\partial \bar F^\alpha} \right \vert_0 \
+ \ \left . \underline{\delta} F^\alpha \frac{\partial  \mathcal{L}_{\rm B}}{\partial F^\alpha} \right \vert_L  
+ \ \left . \underline{\delta} \bar F^\alpha \frac{\partial  \mathcal{L}_{\rm B}}{\partial \bar F^\alpha} \right \vert_L \right \} 
\nonumber \\  & & + \sum_{F=Q,D} \, 
\int d^4x~dy~\left \{ 
\, \partial_M \left ( \underline{\delta} F^\alpha \frac{\partial \mathcal{L}_{\Psi}}{\partial ( \partial_M F^\alpha)} \right ) +
\partial_M \left ( \underline{\delta} \bar F^\alpha \frac{\partial \mathcal{L}_{\Psi}}{\partial ( \partial_M \bar F^\alpha)} \right ) \right \} 
 \  . \nonumber \\
\label{deltabarS-2Yuk}
\end{eqnarray}
Here the four terms in the second line cancel each other since for example the infinite\-simal field variations~\eqref{var-Yuk} lead to,
\begin{equation}
- \left . \underline{\delta} Q^\alpha \frac{\partial  \mathcal{L}_{\rm B}}{\partial Q^\alpha} \right \vert_0 
-  \ \left . \underline{\delta} \bar Q^\alpha \frac{\partial  \mathcal{L}_{\rm B}}{\partial \bar Q^\alpha} \right \vert_0 \
 \, =   \ \left . \frac{1}{2} \bar Q (i\alpha Q)  \right \vert_0 \, + \, \left . \frac{1}{2} (-i\alpha \bar Q) Q  \right \vert_0 \, =  \, 0 \ ,
\label{d-brane-vanishYuk} 
\end{equation}
and the first line (right-hand side) vanishes as for instance the infinite\-simal field variations of type~\eqref{var-Yuk} lead to,
\begin{eqnarray}
 \sum_{F=Q_{L/R},D_{L/R}}  \, \left [ -  \ \left . \underline{\delta} F^\alpha \frac{\partial (Y_5\,  {Q}^\dagger_LHD_R)}{\partial F^\alpha} \right \vert_L 
-  \ \left . \underline{\delta} F^{\dagger\alpha} \frac{\partial (Y_5\,   {Q}^\dagger_LHD_R)}{\partial F^{\dagger\alpha}} \right \vert_L \ \right ] 
= \nonumber \\
  - \ \left .Y_5 \,  {Q}^\dagger_LH(i\alpha D_R)  \right \vert_L \, - \, \left . Y_5\,   (-i\alpha Q^\dagger_L)HD_R  \right \vert_L \, =  \, 0 \ .
\label{d-brane-Yukvanish} 
\end{eqnarray}
Therefore considerations on the vanishing infinitesimal variation~\eqref{deltabarS-2Yuk} over a generic 5D domain $\Omega$, similar as in the free case, lead to 
the local conservation relation for the 5D probability current, 
\begin{equation}
\phantom{0} \partial_M j^M = 0 \, , \  \text{for any $z^M$} , \  \text{with} \ 
j^M \, = \,  \sum_{F=Q,D} \underline{\delta} F^\alpha \frac{\partial \mathcal{L}_{\Psi}}{\partial ( \partial_M F^\alpha)} +
 \underline{\delta} \bar F^\alpha \frac{\partial \mathcal{L}_{\Psi}}{\partial ( \partial_M \bar F^\alpha)}  \  .
\label{currentGENYuk}
\end{equation}

\section{Boundary conditions}
\label{BC-EBC}

In this Appendix, we write down the global boundary condition derived from the initial variation of the action 
$S^{\rm{m}}_{\rm{5D}}$ in Eq.~\eqref{eq:actionNoH}:
\begin{align}
 \delta_{Q^\dagger_L} S\vert_{\rm{b}} \, + \, & \delta_{Q^\dagger_R} S\vert_{\rm{b}} +  \delta_{D^\dagger_L} S\vert_{\rm{b}} + \delta_{D^\dagger_R} S\vert_{\rm{b}} 
+ \delta_{Q_L} S\vert_{\rm{b}} +  \delta_{Q_R} S\vert_{\rm{b}} +  \delta_{D_L} S\vert_{\rm{b}} + \delta_{D_R} S\vert_{\rm{b}} = 0  \, , \   \rm{with,}
\nonumber 
\end{align}
\begin{align}
\delta_{Q^\dagger_L} S^{\rm{m}}_{\rm{5D}} &
\ni \delta_{Q^\dagger_L} S\vert_{\rm{b}} = 
\displaystyle{ \int d^4x \left. \left[ \delta Q^\dagger_L \left( \dfrac{1}{2} Q_R -X D_R \right) \right] \right|_{L} - \int d^4x \ \left. \left( \delta Q^\dagger_L  Q_R \right)  \right|_{0} } \ ,
\nonumber \\
\delta_{Q^\dagger_R} S^{\rm{m}}_{\rm{5D}} & 
\ni \delta_{Q^\dagger_R} S\vert_{\rm{b}} = 
\displaystyle{  \int d^4x \left. \left[ - \delta Q^\dagger_R \left( \dfrac{1}{2} Q_L + X' D_L \right) \right] \right|_{L} } \ ,
\nonumber \\  
\delta_{D^\dagger_L} S^{\rm{m}}_{\rm{5D}} & 
\ni \delta_{D^\dagger_L} S\vert_{\rm{b}} = 
\displaystyle{ \int d^4x  \left. \left[ \delta D^\dagger_L \left( \dfrac{1}{2} D_R -X^{\prime \star} Q_R \right) \right] \right|_{L}  } \ ,
\nonumber \\
\delta_{D^\dagger_R} S^{\rm{m}}_{\rm{5D}} & 
\ni \delta_{D^\dagger_R} S\vert_{\rm{b}} =  
\displaystyle{ \int d^4x \left. \left[ -\delta D^\dagger_R \left( \dfrac{1}{2} D_L + X^{\star} Q_L \right) \right] \right|_{L} + \int d^4x \ \left. \left( \delta D^\dagger_R  D_L \right)  \right|_{0} } \ ,
\nonumber \\
\delta_{Q_L} S^{\rm{m}}_{\rm{5D}} &
\ni \delta_{Q_L} S\vert_{\rm{b}} = 
\displaystyle{ \int d^4x \left. \left[ \left( \dfrac{1}{2} Q^\dagger_R -X^{\star} D^\dagger_R \right)  \delta Q_L \right] \right|_{L} - \int d^4x \ \left. \left( Q^\dagger_R  \delta Q_L \right)  \right|_{0} } \ ,
\nonumber \\
\delta_{Q_R} S^{\rm{m}}_{\rm{5D}} & 
\ni \delta_{Q_R} S\vert_{\rm{b}} = 
\displaystyle{  \int d^4x \left. \left[ - \left( \dfrac{1}{2} Q^\dagger_L + X^{\prime \star} D^\dagger_L \right) \delta Q_R \right] \right|_{L} } \ ,
\nonumber \\ 
\delta_{D_L} S^{\rm{m}}_{\rm{5D}} & 
\ni \delta_{D_L} S\vert_{\rm{b}} = 
\displaystyle{ \int d^4x  \left. \left[ \left( \dfrac{1}{2} D^\dagger_R -X' Q^\dagger_R \right)  \delta D_L \right] \right|_{L}  } \ ,
\nonumber \\
\delta_{D_R} S^{\rm{m}}_{\rm{5D}} & 
\ni \delta_{D_R} S\vert_{\rm{b}} =  
\displaystyle{ \int d^4x \left. \left[ - \left( \dfrac{1}{2} D^\dagger_L + X Q^\dagger_L \right) \delta D_R \right] \right|_{L} + \int d^4x \ \left. \left(   D^\dagger_L \delta D_R \right)  \right|_{0} } \ .
\label{eq:App-noEBC}
\end{align}

\newpage

\bibliographystyle{JHEP}
\bibliography{bibli}
\end{document}